\documentclass[a4paper,fleqn,usenatbib]{mnras}

\usepackage{newtxtext,newtxmath}

\usepackage[T1]{fontenc}
\usepackage{ae,aecompl}

\usepackage{graphicx}	
\usepackage{amsmath}	
\usepackage{amssymb}	
\usepackage{hyperref}

\title[Followup of X-CLASS with GROND]{Cosmology with XMM galaxy clusters: the X-CLASS/GROND catalogue and photometric redshifts}

\author[J. Ridl et al.]{J. Ridl,$^{1}$\thanks{E-mail: jridl@mpe.mpg.de}
N. Clerc,$^{1}$
T. Sadibekova,$^{2}$
L. Faccioli,$^{2}$
F. Pacaud,$^{3}$
J. Greiner,$^{1}$
\newauthor 
T. Kr{\"u}hler,$^{1}$
A. Rau,$^{1}$
M. Salvato,$^{1}$
M.-L. Menzel,$^{1}$
H. Steinle,$^{1}$
P. Wiseman,$^{1}$
\newauthor 
 K. Nandra,$^{1}$
J. Sanders$^{1}$
\\
$^{1}$Max-Planck-Institut für Extraterrestrische Physik, Giessenbachstra\ss e, D-85748 Garching, Germany \\
$^{2}$Service d'Astrophysique AIM, CEA/DSM/IRFU/SAp, CEA Saclay, 91191 Gif-sur-Yvette, France \\
$^{3}$Argelander-Institut für Astronomie, University of Bonn, Auf dem Hügel 71, 53121 Bonn, Germany \\
}

\date{Accepted XXX. Received YYY; in original form ZZZ}

\pubyear{2017}

\begin{document}
\label{firstpage}
\pagerange{\pageref{firstpage}--\pageref{lastpage}}
\maketitle

\begin{abstract}
The XMM Cluster Archive Super Survey (X-CLASS) is a serendipitously-detected X-ray-selected sample of 845 galaxy clusters based on 2774 XMM archival observations and covering approximately 90 deg$^2$ spread across the high-Galactic latitude ($|b|>20\degr$) sky. The primary goal of this survey is to produce a well-selected sample of galaxy clusters on which cosmological analyses can be performed. This article presents the photometric redshift followup of a high signal-to-noise subset of 266 of these clusters with declination $\delta<+20\degr$ with GROND, a seven channel ($grizJHK$) simultaneous imager on the MPG 2.2m telescope at the ESO La Silla Observatory. We use a newly developed technique based on the red sequence colour-redshift relation, enhanced with information coming from the X-ray detection to provide photometric redshifts for this sample. We determine photometric redshifts for 236 clusters, finding a median redshift of $z=0.39$ with an accuracy of $\Delta z = 0.02 (1+z)$ when compared to a sample of 76 spectroscopically confirmed clusters. We also compute X-ray luminosities for the entire sample and find a median bolometric luminosity of $7.2\times10^{43} \mathrm{erg\ s^{-1}}$ and a median temperature 2.9 keV. We compare our results to the XMM-XCS and XMM-XXL surveys, finding good agreement in both samples. 
The X-CLASS catalogue is available online at \url{http://xmm-lss.in2p3.fr:8080/l4sdb/}.
\end{abstract}



\section{Introduction}
\label{sec:introduction}
A significant goal of modern astronomy is to provide observations capable of testing the current cosmological paradigm, where the energy density of the Universe is dominated by the cosmological constant, $\Lambda$, and cold dark matter ($\Lambda$CDM). Since the number density of galaxy clusters as a function of mass and redshift depends strongly on various cosmological parameters such as $\Omega_M$, $\sigma_8$ and the physical properties of dark energy, observations of clusters provide a powerful probe of the underlying cosmological model. The parameters $\Omega_M$ and $\sigma_8$ can be well constrained given a sufficiently large sample of low redshift clusters, spanning a wide range of masses. On the other hand, a sample spanning a wide range of masses and redshifts is necessary to place competitive constraints on evolutionary parameters such as the dark energy equation of state \citep{Vikhlinin2009a}. Such a sample can also be used to study the evolution of various cluster scaling relations, such as the X-ray luminosity or temperature to total cluster mass ($L_X-M$ and $T_X - M$). Of crucial importance to any attempt to use clusters for cosmological studies is an intricate knowledge of the sample selection function and how it is related to the underlying cluster distribution, predicted by cosmological simulations. For a comprehensive review on clusters as cosmological probes, see \citet{Allen2011} and the references therein.  

The most obvious way in which galaxy clusters can be identified and selected is as an over-density in the spatial distribution of galaxies, particularly in optical and near-infrared (NIR) wavelengths \citep[e.g][]{Abell1958, Gladders2000a, Rykoff2014}. Such samples are however difficult to characterise due to the lack of highly constrained scaling relations for moving from directly observable quantities, such as the cluster richness to the total halo mass. Further, they are generally more contaminated due to projection effects than other methods  e.g. redMaPPer reports an incidence of contamination of $\sim 5\%$ \citep{Rykoff2014}. A significant advantage of optical/NIR cluster detection algorithms is that they typically produce an estimate of the cluster redshift, thanks to the well studied and constrained colour-redshift relation of passively evolving galaxies, which make up the cluster red sequence  \citep{Baum1959}. 

The baryonic component of galaxy clusters typically takes the form of a hot intracluster gas which is detected either directly through its X-ray emission, or indirectly via the Sunyaev-Zel'dovich (SZ) decrement \citep{Sunyaev1970}. Methods taking advantage of this are less likely to be affected by projection effects but do not readily provide any redshift information in general. However, given a robust estimate of the redshift from followup optical observations, the intra-cluster gas provides a ready proxy of the total halo mass and is thus an excellent probe of the halo mass function. 

It is thus clearly optimal to perform studies of galaxy clusters over a wide range of wavelengths to fully exploit all the available information. Many studies have followed this philosophy, whereby clusters are detected through their X-ray emission and then followed-up with ground or space-based optical and NIR observations to confirm the cluster candidate and to obtain the redshifts needed for their physical characterisation. Examples of these include wide-field surveys with ROSAT \citep{Vikhlinin1998, Boehringer2000}, medium-field observations with XMM e.g. XMM-LSS \citep{Pierre2007, Pacaud2007, Clerc2014}, XMM-XXL \citep{Pierre2016, Pacaud2016} and  XMM-BCS \citep{vsuhada2012xmm} and narrow surveys such as the COSMOS field with Chandra \citep{Scoville2007} or XMM \citep{Finoguenov2007}. Additionally, the vast number of PI observations with XMM and Chandra provides an abundance of exploitable data in which serendipitous cluster searches can be performed with Chandra \citep[ChaMP,][]{Barkhouse2006} and with XMM e.g. XCS \citep{Romer2001,Lloyd-Davies2011,Mehrtens2012} and X-CLASS \citep{Clerc2012, Sadibekova2014}. The sample presented in this paper, X-CLASS lies in the middle ground between the XCS and XXL surveys in that the pointings are distributed across the entire extragalactic sky and yet the detection of clusters take place on pointings with homogeneous exposure times. 

A wide variety of techniques and methods have been used to identify cluster of galaxies in large, wide-area optical surveys, making use of various well known properties of clusters. One of the well studied features of galaxy clusters that is commonly used for their detection is the presence of the cluster red-sequence which takes advantage of the colour-magnitude relation (CMR) of early-type galaxies due to the 4000 Å break in their rest frame, \citep[e.g.][]{Gladders2000a}. The algorithm of maxBCG \citep{Koester2007}, also takes advantage of the existence of a unique brightest cluster galaxy (BCG) which lies on the red sequence. More recently, redMaPPer \citep{Rykoff2014} and WHL \citep{wen2012,wen2015} have provided optimised methods for the detection of optical clusters and accurate determination of the redshift and richness. For the photometric redshifts derived in this paper we extend the red sequence method to take advantage of the prior knowledge that we obtain from the X-ray detection of the cluster, namely the position of the cluster centre and the extent of the X-ray emission.

This paper is structured as follows. In Section \ref{sec:xclass} we present a summary the XMM Cluster Archive Super Survey (X-CLASS) focusing on the source detection and sample selection. We then describe our optical and near-infrared followup program with GROND in Section \ref{sec:grond} and discuss the redshift determination in Section \ref{sec:redshifts}. The measurement of the X-ray properties of our sample is discussed in Section \ref{sec:x-ray} and the results and discussion of interesting cases are presented in Section \ref{sec:results} and Section \ref{sec:discussion} respectively. The cosmological analysis, based on the the forward-modelling approach of \citet{Clerc2012a}, will be presented in a companion paper (Ridl et al., in prep).

Throughout, we assume a $\Lambda$CDM cosmological model relying on the parameters calculated by \citet{hinshaw2013wmap}, in particular with $\Omega_M=0.28$, $\Omega_\Lambda=0.72$ and $H_0=70\ \mathrm{km\ s^{-1} Mpc^{-1}}$.

\section{The XMM Cluster Archive Super Survey}
\label{sec:xclass}
X-CLASS is a serendipitous search for galaxy clusters in archival observations from the XMM-Newton observatory, with the main objective of producing a well defined-cluster sample suitable for cosmological studies. The data were processed utilising the procedures of the XMM-LSS collaboration \citep[][Faccioli et al. in prep]{Pacaud2006}, and the construction of the X-CLASS catalogue is described in \citep{Clerc2012}. We summarise the key points here.
\subsection{Selection of XMM pointings and cluster detection}
The following constraints were taken into account when selecting observations from the XMM Science Archive system from publicly available data, as of 26 May 2010, for analysis. In order to reduce the impact of galactic foregrounds, we selected only pointings centred at Galactic latitudes ${|b|\geq 20}$ deg and located (5 deg / 2 deg) from (Magellanic Clouds / M31). Further, we required that the exposure time (given by the duration in the XMM archive) was greater than 5 ks and that all three detectors (MOS1, MOS2 and PN) were in imaging mode, with at least one being in Full Frame mode.

\subsubsection{Processing of data}
The calibrated event lists are first filtered from proton and solar flares resulting in a \textit{good time interval} (GTI) which is used to proceed with the analysis. The overall quality of each observation was then visually inspected and some observations discarded. 

Since clusters detected with XMM exposure times of ~10-20 ks form a highly relevant population for cosmological studies \citep{Pierre2007,Pierre2011} and the implementation of a survey selection function is simplified when working with a survey consisting of homogeneous exposure times, new pointings are built from the original exposures so that each pointing is cut to either a 10 or 20 ks exposure time on the three detectors, after correcting for background flares. Once observations where one or more of the detectors had a GTI of less than 10 ks were removed, the total number of pointings from which sources are detected is 2409, giving a total exposure time of the survey of 24 Ms out of a possible 40 Ms of good-time-intervals (GTI) available. 

\subsubsection{X-ray source detection}
\begin{figure*}
  \centering  
  \includegraphics[width=1.99\columnwidth]{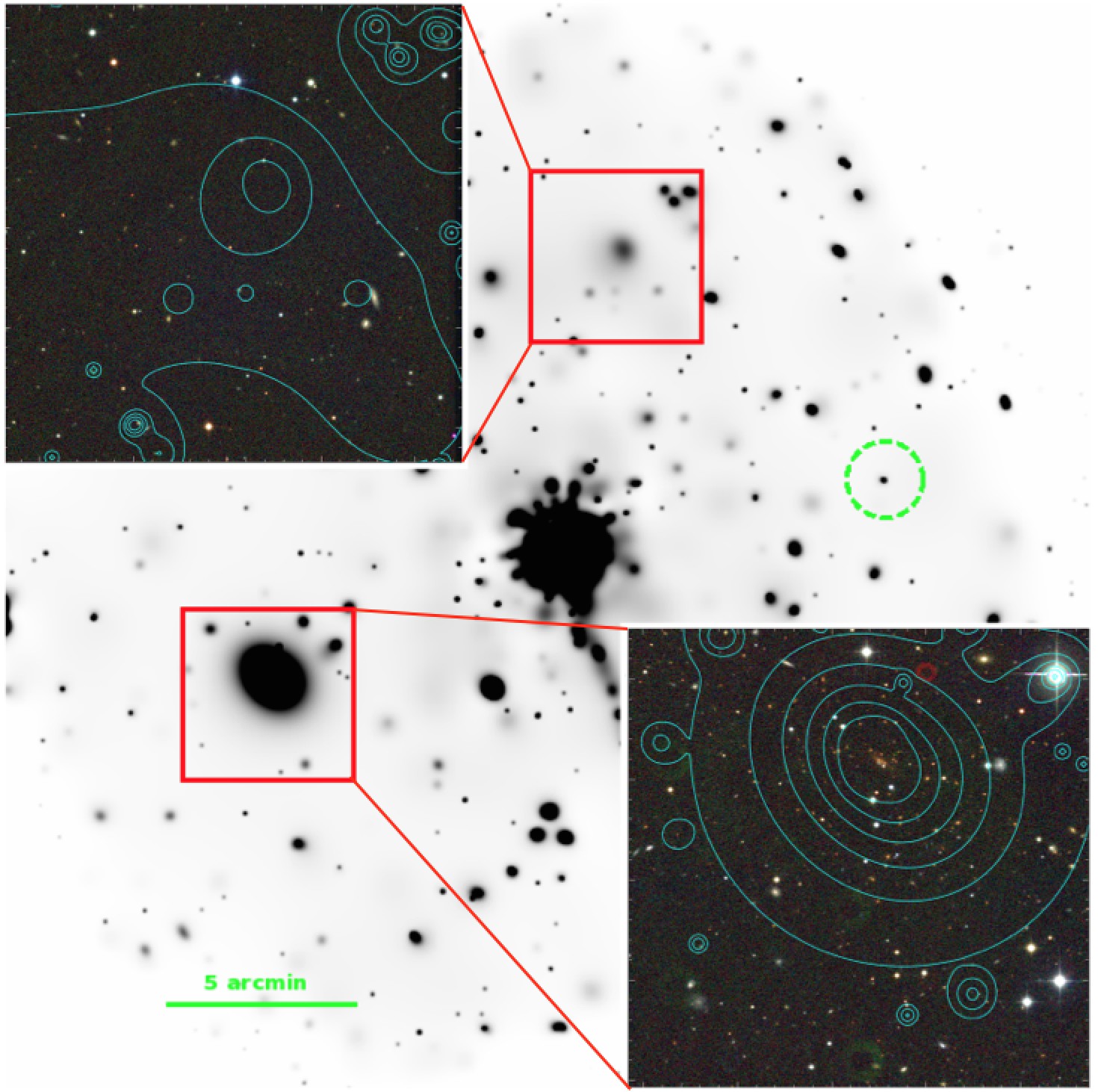}
  \caption{Wavelet filtered (M1$+$M2$+$PN) image, ObsID: 0555020201\_20ks. Red boxes show the locations of two serendipitously detected C1$^+$ clusters, X-CLASS 2305  ($z=0.62$) and X-CLASS 2304 (distant candidate; see discussion in Section \ref{sub:distant}) along with the GROND $gri$ image Cyan contours represent the X-ray distribution. The PI target, RBS 1055 is located near the centre of the pointing. For comparison, a point source is indicated by the dashed-green circle.}
  \label{fig:waveletFilteredImage}
\end{figure*}

The detection of sources is performed on a co-added image of the three EPIC detectors in the [0.5 - 2] keV range of each of the three EPIC detectors. The source extraction tool \texttt{SExtractor} \citep{Bertin1996} is run on a wavelet-filtered \citep[\texttt{mr\textunderscore filter,}][]{starck1998image,valtchanov2001comparison}, co-added image and only sources detected within 13 arcmin of the pointing centre are considered for further analysis. A maximum likelihood profile fitting procedure \citep[\texttt{XAMIN},][]{Pacaud2006} further characterises the detected source as being either point-like or extended, i.e. a $\beta$-model convolved with the PSF.  A set of parameters characterising each of the detected sources is also provided, including the angular extent (EXT), which defines the apparent core radius of the best fit $\beta$-model and the likelihood that the emission is extended (EXT\textunderscore LIKE). Flux measurements are performed on the `full exposure' pointings, after removing periods of high-background, containing the maximal available GTI for each observation, enhancing the signal-to-noise. Figure \ref{fig:waveletFilteredImage} illustrates a wavelet-filtered XMM image containing 2 detected clusters and GROND $gri$-images of the cluster positions overlaid. 

\subsection{Catalogue construction and selection of the cosmological sub-sample}
Following \citet{Pacaud2006} a catalogue is built by selecting extended sources within 13\arcmin of the centre of the parent pointing with EXT$>5\arcsec$ and EXT\_LIKE > 33. Such sources are denoted `C1'. This selection results in a low ($<5\%$) level of contamination by incorrectly classified point-sources. There are a variety of astronomical objects present in the observations and to accurately remove large nearby clusters, nearby galaxies, planets and unresolved double or saturated point-sources, human intervention is necessary. After removal of duplicate detections, all candidate clusters were screened by at least two independent astronomers based on optical data from the Digitized Sky Survey (DSS) POSS-II with the X-ray contours overlaid. Each astronomer awarded a `quality' flag to the detection and a final decision was made by a moderator based upon the evaluators' comments. In addition to a decision being made on the nature of the source, the DSS imaging was also used to give a rough estimate of the possible redshift range of the clusters, dividing them into categories of $0<z<0.3$ and $z>0.3$. As of Aug. 2010, the catalogue contains 845 C1 clusters. 

\subsubsection{The cosmological sample}
The primary goal of this paper is to describe a catalogue for use in cosmological calculations, extending the previous CR-HR (\textit{count rate - hardness ratio}) analysis with the addition of cluster redshift information i.e. z-CR-HR \citep{Clerc2012a}. For this purpose, a high signal-to-noise ratio subsample is selected according to the following criteria:
\begin{enumerate}
\item The data set was selected by removing pointings with high background; with one or more detectors not being in full frame mode; and those centred on luminous nearby clusters. This results in the total area surveyed for use in the cosmological fits of 1992 pointings. 
\item A more pure sub-class of galaxy clusters  with EXT\_LIKE > 40, denoted by `C1$^+$' was selected and included in the catalogue. 
\item A final cut was made in terms of the measured X-ray properties of the sources namely CR, as the count-rate measured in the [0.5-2] keV range and HR, the ratio between the [1-2] keV and [0.5-1] keV count-rates. We summarise these measurements in Section \ref{sub:gc}. Only clusters with 0.009 $<$ CR $< 0.5\ \mathrm{cts\ s^{-1}}$  and 0.05 $<$ HR $<$ 2 were included in the final cosmological subsample consisting of 461 clusters. 
\end{enumerate} 
We account for the C1+ cluster selection by modelling the cluster population in the observable domain. Unobserved objects are filtered out by using the observable-based selection function derived from realistic XMM observations (see e.g. \citet{Pacaud2006} for the definition of C1, \citet{Clerc2012} for the application to the CR-HR modelling, \citet{Pacaud2016} for the $dn/dz$ modelling, \citet{Giles2016} for the modelling of the luminosity-temperature $L-T$ relation, and references therein).

The optical and near-IR followup of a Southern ($\delta<20\degr$) subset of 266 of these clusters, visible from the ESO La Silla observatory in Chile forms the basis of the rest of this paper. 

\section{Optical and near-IR followup with GROND}
\label{sec:grond}
One of the main goals of this work is to provide photometric redshifts for X-ray selected galaxy clusters in order to perform a cosmological analysis. 

To achieve this, an extensive followup campaign with the \textbf{G}amma-\textbf{R}ay Burst \textbf{O}ptical and \textbf{N}ear-Infrared \textbf{D}etector (GROND) \citep{Greiner2008} on the MPG 2.2m telescope at the ESO La Silla Observatory was undertaken. The observations were performed over 6 observing periods (ESO periods P91-P96) and 77 nights between April 2013 and February 2016. More information detailing the observations are presented in Appendix \ref{app:observing}. GROND is a 7-channel imager, allowing for simultaneous imaging in the Sloan $g',r',i',z'$ and near-infrared $JHK$ bands. It was primarily designed to provide rapid multi-wavelength observations of gamma-ray burst afterglows e.g. \citep{greiner2009grb, gkk11, gmk15}. For the remainder of this paper GROND optical filters will be expressed as $g$, $r$, $i$ and $z$

\begin{figure}
\centering 
\includegraphics[width=0.99\columnwidth]{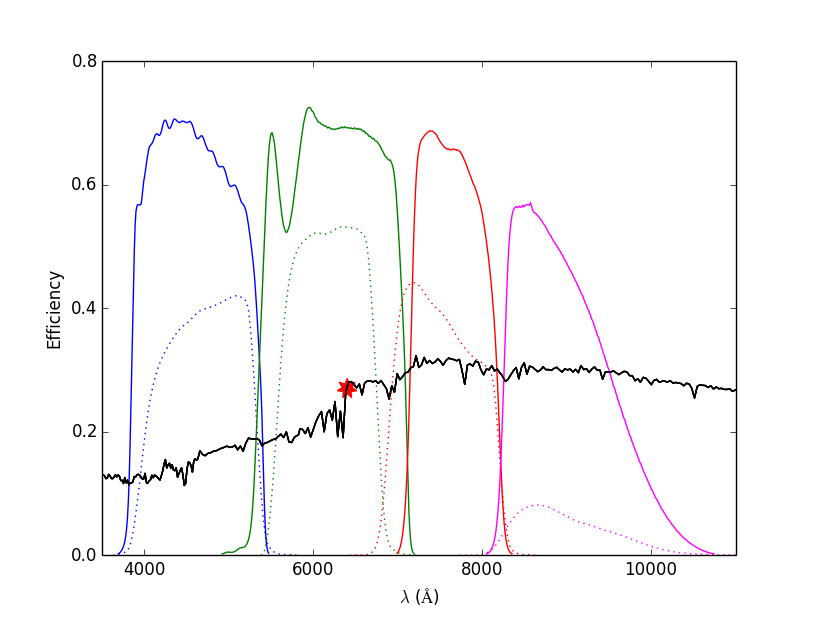}
  \caption{The efficiency of GROND (solid) and for comparison, SDSS (dotted) filters is shown as a function of wavelength. The narrow width of the GROND $i$-band compared to SDSS is clearly visible. Also plotted is an arbitrarily scaled spectral energy distribution (SED) of a early type galaxy at redshift $z=0.6$. The $4000 \AA$ break, the key feature used for determination of the redshift of the cluster is marked with a star.}
  \label{fig:filters}
\end{figure}

Incoming light is split into different photometric bands by making use of dichroics and the design is such that the optical transmission functions are almost identical to those of the Sloan $g',r',i',z'$ filter system. The exception to this is the $i$-band which, due to the overlap between the Sloan $r',i',z'$ bands, is slightly narrower, in favour of standard-width $r$ and $z$ bands. Each of the optical CCDs provides a field-of-view of $5.4 \arcmin \times 5.4 \arcmin$ with a pixel scale of $0.158\arcsec$ pixel$^{-1}$. The optical filter transmission curves are shown in Figure \ref{fig:filters}. The NIR part of GROND is a focal reducer system and provides a $10 \times 10$ arcmin$^2$ field-of-view with a pixel scale of $0.60''$ pixel$^{-1}$. The $K$ channel additionally includes a flip mirror for dithering purposes. For the remainder of this work we consider only the optical channels since they span the 4000 \AA\ break, which is the most redshift-constraining feature for early-type galaxies, over the redshift range in which were are interested. A comprehensive description of the instrument is given in \citet{Greiner2008}. 

\subsection{Operation of GROND}
\label{sub:grond}
In operating a 7-channel simultaneous imager there are several observational constraints that need to be taken into account when preparing observation blocks. The optical and near-IR systems require a different number of exposures at each telescope dithering position and integration times for each of them should be set to optimise the exposure time also taking the differing read-out time of the detectors into account. 

We determined that four telescope dithering positions with a single optical exposure at each would be sufficient for our observations and there are a number of predefined OB types, named for the total integration in the $K$-band namely 4-, 8-, 20-, 40-minute OBs which satisfy this constraint. Two read-out modes for the optical CCDs are available namely `fast' and `slow'. The exposure times for these OBs are given in Table \ref{tab:obs}. 

\begin{table}
\begin{center}
 \caption{Total exposure times of the predefined GROND observing blocks in the optical ($griz$) and near-IR ($JHK$) channels used in this study. Execution times are approximate and include telescope slewing.}
    \begin{tabular}{|c|c|c|c|c|}   
	OB type& $griz$ & $JHK$  & Read-out & Execution time\\
	 & ($s$) & ($s$)& & (min)\\
	  \hline
	4min4TD & 141.6  & 240 & Slow & 10\\
	4min4TD &   264.0 & 240 & Fast & 10\\
	8min4TD & 459.6 & 480 & Slow & 15\\
	8min4TD & 579.6 & 480 & Fast & 15\\
	20min4TD & 1476.0 & 1200 & Slow & 30\\
	20min4TD & 1596.0 & 1200 & Fast & 30\\
 
	 \hline
	 \label{tab:obs}
    \end{tabular}
\end{center}
\end{table}
Initial pathfinding observations indicated that sufficient depth is obtained for clusters of $z<0.3$ and $z>0.3$ with the 8min4TD and 20min4TD OBs respectively. Standard fields used for photometric calibration are observed with 4min4TD OB. All observations were initially carried out in slow read-out mode until November 2015 when a technical issue necessitated a change to fast read-out mode with its somewhat higher read-out noise. 

\subsection{Data reduction and image combination}
\label{sub:data_reduction}

Preliminary reduction of the data was performed for each OB using the methods of \citet{Yoldas2008} and \citet{Kruhler2008}. This pipeline is based on the standard tools of \textit{IRAF/PyRAF} and performs bias and dark current subtraction, flat-fielding and defringing along with providing  astrometrised co-added images and a photometric measurements idealised for point sources for each channel. The main steps are summarised here.

A number of standard bias and dark frames were recorded directly at the end of each observing night with a wide range of exposure times for the dark frames. Master bias and dark frames were then produced by combining the individual exposures. Flat field observations were performed during twilight either in the evening preceding or in the morning following the observing program for each night when the conditions allowed. The GROND flat fields have been shown to be consistent over a number of nights so if weather conditions did not allow for the successful acquisition of suitable flat field, those recorded on a nearby night were used. Great care was taken to ensure that all seven simultaneously observed skyflats were suitably exposed and that the sky was bright enough to obtain a statistically robust flat-field without saturating the detectors. The removal of the bias and dark current, and the correction for the pixel-to-pixel variations on the CCD were performed simultaneously with the IRAF tool \texttt{quadred.ccdproc}.

As with most optical instruments, the $i$- and $z$-bands of GROND are affected by fringing effects. A master fringe pattern for each OB and each of these bands is created by combining those generated for each of the four individual exposures. This master pattern is then subtracted from each frame individually (IRAF \texttt{rmfringe}) before they are combined into the final co-added image (coadd). The individual exposures are combined to form a single coadd for each filter using IRAF \texttt{imcombine}. Finally, the sky background is calculated from each of these coadds with the sources masked out and subtracted from the image. 

\subsection{Astrometry and Photometry}
\label{sub:astrophot}
An astrometric solution was accomplished through the matching of stars in common with SDSS DR7 \citep{abazajian2009_sdss} when available or the USNO-A2.0 catalogue \citep{Monet1998_usno} where the observations fell outside the footprint of SDSS for the optical bands and the 2MASS catalogue \citep{skrutskie2006_2mass} for the near-IR bands and making use of the IRAF tool \texttt{xyxymatch}. The astrometric solution was refined by making use of the publicly available software \texttt{SCAMP} \citep{bertin2006_scamp} and the coadded images in the respective bands mapped to a common pixel grid with a scale of 0.158\arcsec $\mathrm{pixel^{-1}}$ with the use of \texttt{SWARP} \citep{bertin2002_swarp}, a publicly available software that performs the resampling and co-addition of FITS images. 

A general model for the PSF across the field-of-view was constructed from bright, unsaturated stars and making use of the publicly available software \texttt{PSFEx} \citep{bertin2011_psfex}, for which the various parameters were tuned to optimise the accuracy.

Source detection and photometric measurements were performed using \texttt{SExtractor} \citep{Bertin1996}, operating dual-mode with a \texttt{SWARP} $riz$-coadd as the detection image. This forces the photometric measurements to be performed in the same extraction radius for each channel. In this first step, the typical photometric zeropoints for each of the GROND channels were assumed, together with the necessary corrections for exposure time and atmospheric absorption, quantified by airmass. Where possible, the resulting photometric catalogue was cross-matched with SDSS DR7 photometric catalogue with a 1 arcsec matching radius, and non-saturated and unblended stars selected in order to calibrate the zeropoints. The final zeropoints in each channel were then determined by comparing PSF magnitudes (MAG\_PSF from \texttt{SExtractor}) in the two catalogues with the SDSS photometry corrected using the conversion relations given at \url{http://www.mpe.mpg.de/~jcg/GROND/}.

Given that the majority of our sample lies to the south of the SDSS footprint, it was not possible to calibrate the individual zeropoints for each observation. For these fields we attempted to make use of stellar-locus regression methods to obtain a colour-colour calibration but these efforts were typically hampered by an insufficient number of stars lying in the GROND field-of-view. Ultimately, it was decided that the most reliable way to achieve a homogeneous photometric calibration would be to determine a `master' calibration for each observing night. This was accomplished by averaging the zeropoint corrections obtained from the comparison with SDSS for all possible fields, including science and standard star observations. The standard deviation $\sigma$ of the zeropoints was also calculated and fields with zeropoints more than 3$\sigma$ from the average were excluded and the average recalculated. Once this master calibration was in-hand, corrections for extinction due to airmass were applied to each observation independently. The master zeropoints for each observing night are displayed in Figure \ref{fig:conditions}. This plot provides a useful indication of the photometric quality of each night, as discussed in Section \ref{sub:data_quality}.

 Corrections for Galactic extinction were then applied to the GROND object magnitudes based on the dust maps of \citet{schlegal1998_dust}.\footnote{Making use of the Python package Astroquery.}
 
Star-galaxy separation was accomplished by selecting objects based on the \texttt{SExtractor} parameters CLASS\_STAR and SPREAD\_MODEL for the $r$-band and only objects with FLAG $= 0$ in all bands included in the final catalogue. Kron magnitudes, MAG\_AUTO were chosen for the total galaxy magnitudes and for the determination of galaxy colours. 

\subsection{Data quality control}
\label{sub:data_quality}
\begin{figure*}
  \centering  
  \includegraphics[width=2.0\columnwidth]{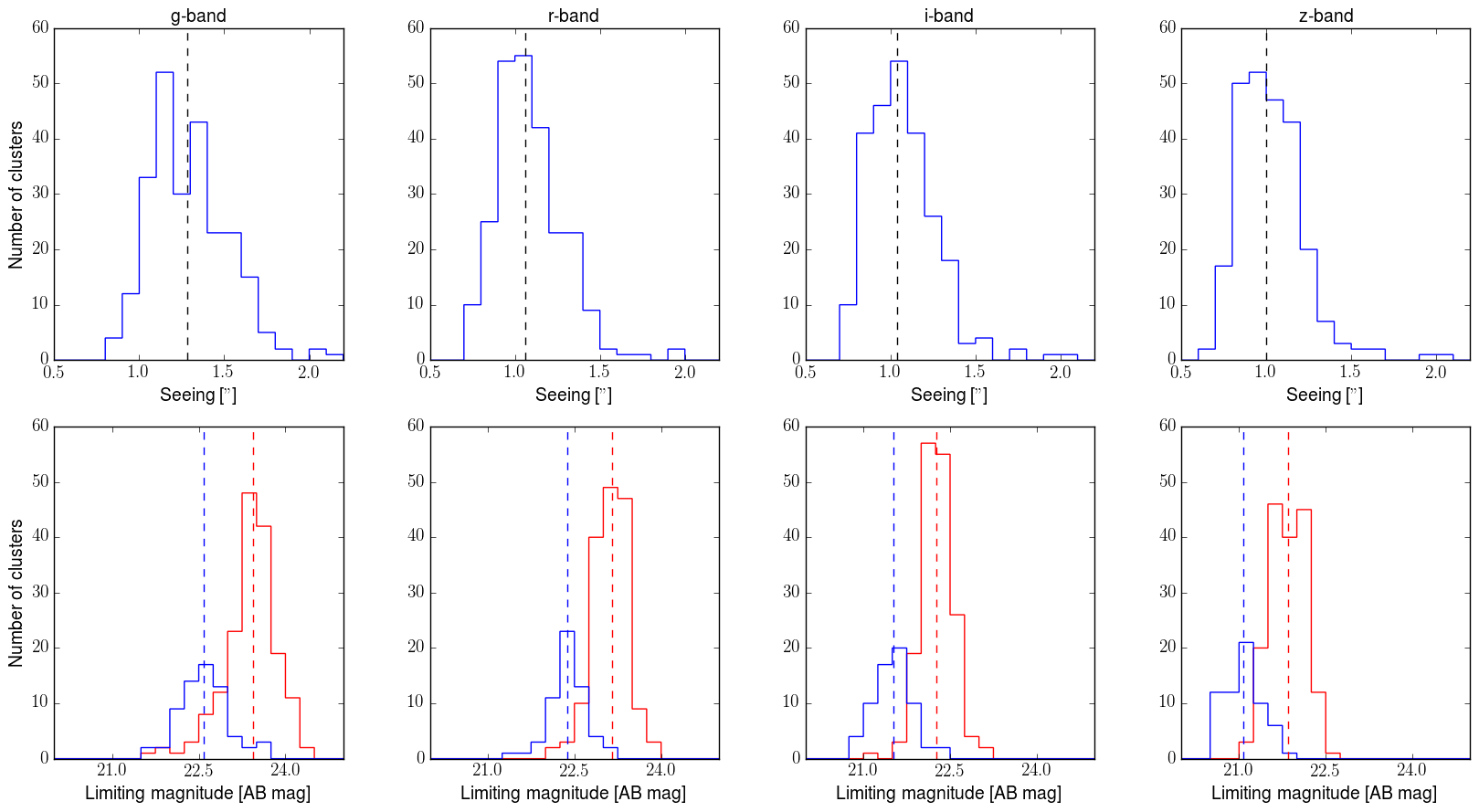}
  \caption{The upper panels describe the distributions of the measured seeing and the lower panels describe the $10\sigma$ point-source limiting magnitudes for 8min4TD (blue) and 20min4TD (red) for each of the $g,r,i,z$-bands. The median values are indicated by the vertical dashed lines. }
  \label{fig:seeing}
\end{figure*}

All observations were inspected visually in terms of the astrometric solution and photometric calibration. In cases where a single galaxy cluster was observed on more than one occasion, the best observation was selected based on seeing, background and limiting magnitude. The stability of the photometric zeropoint calibration for each particular night was also taken into account. The average data quality in terms of seeing and limiting magnitude is summarised in Table \ref{tab:seeing} and Figure \ref{fig:seeing}. 
\begin{table}
\begin{center}
 \caption{The median seeing and $10\sigma$-limiting magnitude in each of the four optical channels and for each of the chosen observing blocks. The limiting magnitudes are determined by the magnitude at which the signal-to-noise for an extended source reaches 10.}
    \begin{tabular}{|c|c|c|c|}   
    	Channel & Seeing & 8min4TD & 20min4TD \\
	 & [\arcsec] & [mag AB] & [mag AB]\\
	  \hline
	$g$ & 1.28 & 22.59 & 23.44 \\
	$r$ & 1.06 & 22.38 & 23.15\\
	$i$ & 1.04 & 21.52 & 22.25\\
      	$z$ & 1.00 & 21.07 & 21.84\\
	 \hline
	 \label{tab:seeing}
    \end{tabular}
    
\end{center}
\end{table}
\begin{figure*}
  \centering  
  \includegraphics[width=2.0\columnwidth]{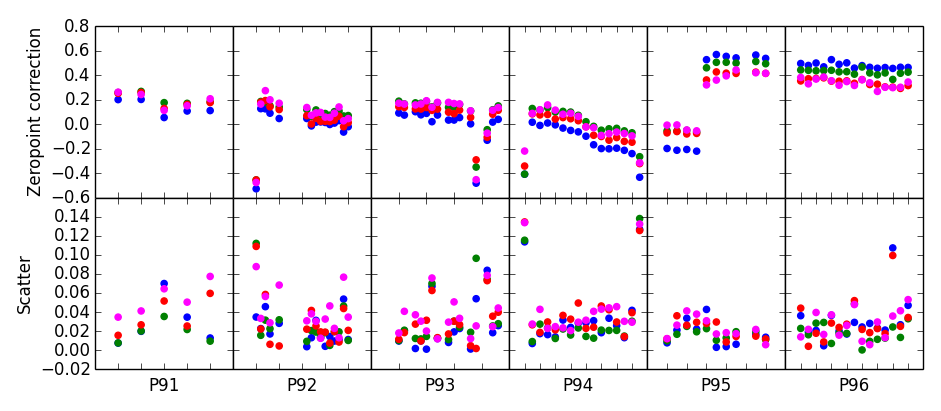}
  \caption{The evolution of the GROND photometric zeropoints in each of the optical bands ($g$: blue, $r$: green, $i$: red, $z$: magenta) over the course of the observing period from ESO periods 91 (starting April 2013) to 96 (ending February 2016). Each point represents the median zeropoint correction for a given observing night, measured from all fields overlapping with the SDSS footprint and after taking into account corrections for atmosphere extinction and differing exposure times. The top panel indicates the zeropoint correction and the bottom panel a measure of the RMS scatter across all measured fields on a given night. }
  \label{fig:conditions}
\end{figure*}
It is interesting to note the evolution of the photometric zeropoints in each channel over the course of the observations, as illustrated in Figure \ref{fig:conditions}. Over the first 4 ESO periods (actual dates of observations are given in Appendix \ref{app:observing}) of observations (P91-P94) we notice a gradual decline in the zeropoints in all channels. This is predominantly due to the collection of dust and the gradual deterioration of the primary mirror of the telescope. During P95, the primary mirror was cleaned and recoated providing a large increase in the photometric depth of the instrument, most notably in the $g$-band where an improvement of $~0.7$ magnitudes is noted. Points significantly below the gradual trend in the zeropoint evolution and those where the scatter is higher than average give a good indication that the night was not photometric and that the calibration can not be trusted. Fields observed on these nights were typically reobserved on nights with higher photometric quality. 

\begin{figure*}
  \centering  
  \includegraphics[width=0.98\columnwidth]{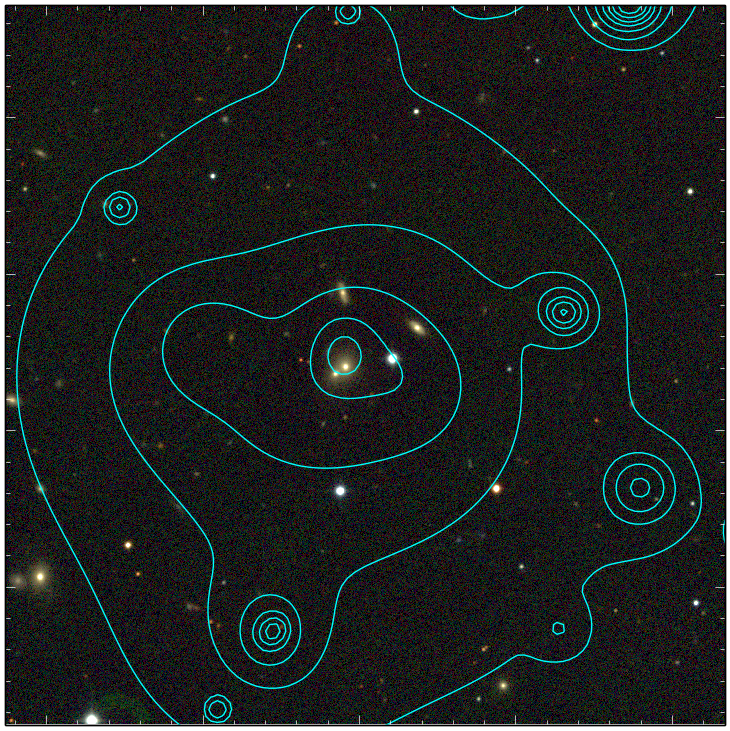}
  \includegraphics[width=0.98\columnwidth]{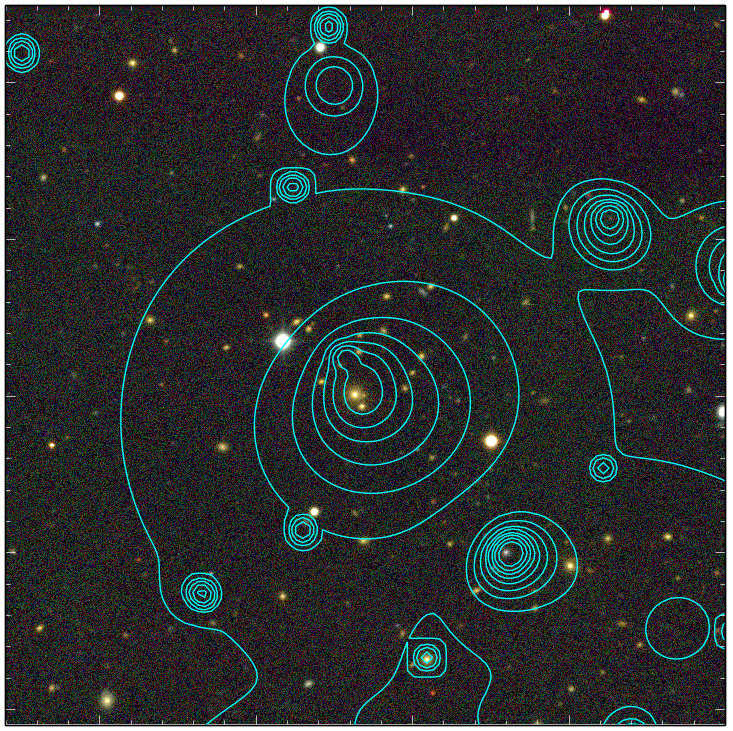}
  \includegraphics[width=0.98\columnwidth]{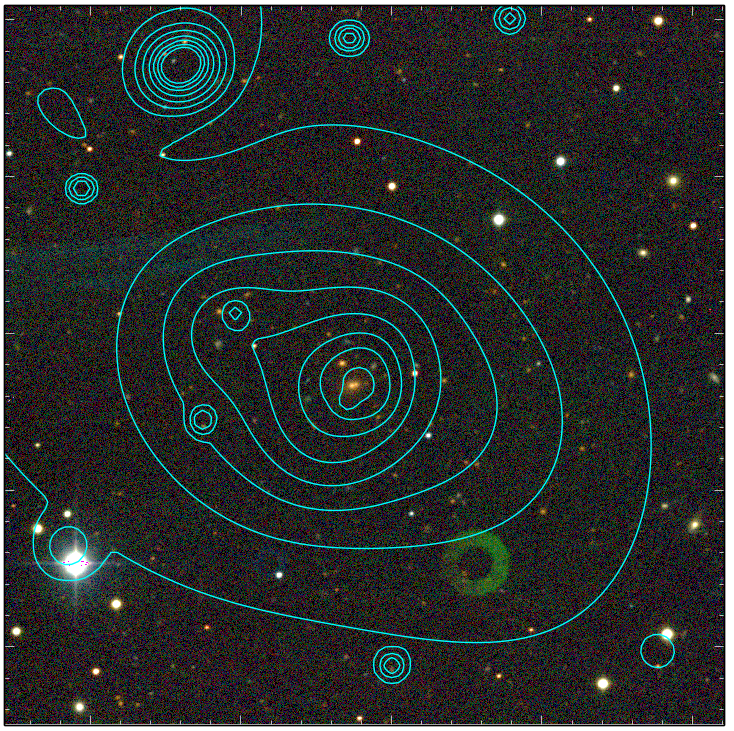} 
  \includegraphics[width=0.98\columnwidth]{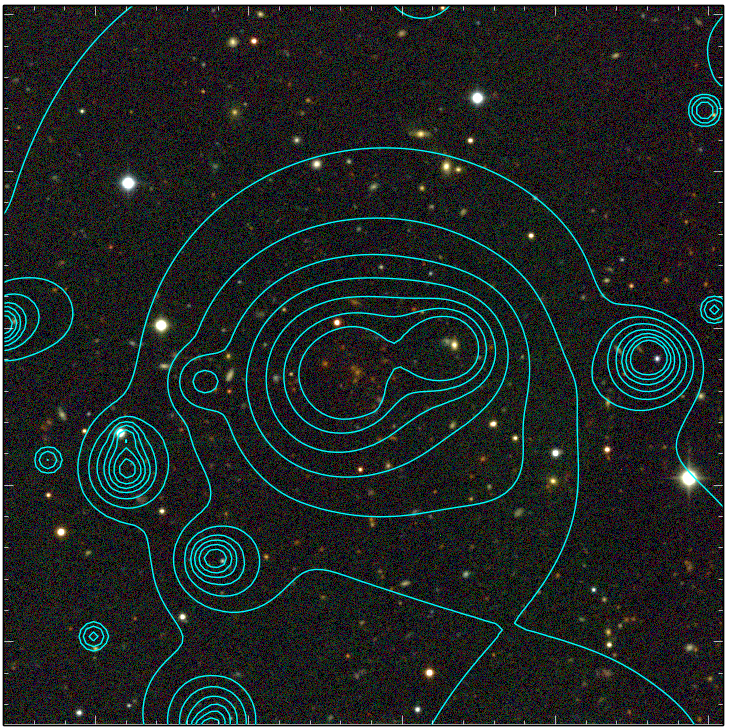}

  \caption{A selection of $g'r'i'$ three-colour composite images for optically confirmed clusters over a range of redshifts. All images are 4.5 $\times$ 4.5 arcmin and the cyan contours are drawn from the wavelet-filtered X-ray images in the [0.5-2] keV range. This compilation shows from left to right and top to bottom: X-CLASS 2162 ($z_{\mathrm{spec}}=0.12$, $z_{\mathrm{phot}}=0.12$);  X-CLASS 40 ($z_{\mathrm{spec}}=0.33$, $z_{\mathrm{phot}}=0.32$); X-CLASS 459 ($z_{\mathrm{spec}}=0.55$, $z_{\mathrm{phot}}=0.54$); X-CLASS 505 ($z_{\mathrm{spec}}=0.79$, $z_{\mathrm{phot}}=0.81$).}
  \label{fig:clusterImages}
\end{figure*}

\section{Redshift measurements}
\label{sec:redshifts}

\subsection{Archival redshifts}
\label{sub:archival}
A comprehensive search for archival redshift information making use of the NED database was undertaken. Where counterparts to our clusters were found, a redshift was allocated to the cluster along with a flag indicating the redshift status. The criteria for each of these status flags are are follows:
\begin{enumerate}
\item \textit{Confirmed}: Abell \citep{Abell1958}, Planck \citep{planck2014planck,ade2015planck}, SPT \citep{bleem2015spt}, XCS-DR1\citep{Mehrtens2012} or other published clusters with spectroscopic redshifts are available; there are at least 3 similar spectroscopic redshifts within 3\arcmin; or there is an obvious BCG with a spectroscopic redshift and many similar photometric redshifts within 3\arcmin.
\item \textit{Photometric}: There is a photometric redshift available for a cluster matched in the XCS-DR1 or elsewhere in literature; or the X-ray position is coincident with a redMaPPer candidate.
\item \textit{Tentative}: There is at least 1, but fewer than 3 similar spectroscopic redshifts.
\end{enumerate}
In total, we find that 88 clusters are already spectroscopically confirmed and a further 66 have a photometric redshift. We find that 25 clusters are allocated the redshift flag `tentative,' but these should be treated with caution and the redshift should by no means be considered to be definitive. 

\subsection{The GROND cluster photometric redshift tool}
\label{sub:icrt}
Observing galaxy clusters with GROND in multiple bands simultaneously has several advantages, since a single pointing results in a multi-chromatic data set obtained under identical atmospheric conditions. This implies that non-photometric conditions have a minimised effect on galaxy colours compared to data taken under varying conditions. The relatively small field-of-view however does introduce some challenges to any attempt to determine cluster photometric redshifts for two main reasons. Firstly, as discussed in Section \ref{sub:astrophot}, it is difficult to obtain an absolute photometric zeropoint calibration due to the lack of stars present in extragalactic fields. Secondly, in most cases the entire field-of-view is taken up by the galaxy cluster itself and it is thus not feasible to obtain an estimate of the local background distribution of galaxies. This makes it difficult to perform an analysis similar to that of redMaPPer \citep{Rykoff2014} or other similar techniques which require secure knowledge of the background galaxy distribution to which any over-densities can be compared. We thus developed our own algorithm to calculate cluster photometric redshifts based on the cluster red sequence colour-redshift technique with the addition of extra information obtained from the X-ray detection of the cluster. 

\subsubsection{Red sequence colour-redshift relation}
\begin{figure}
  \centering  
  \includegraphics[width=0.99\columnwidth]{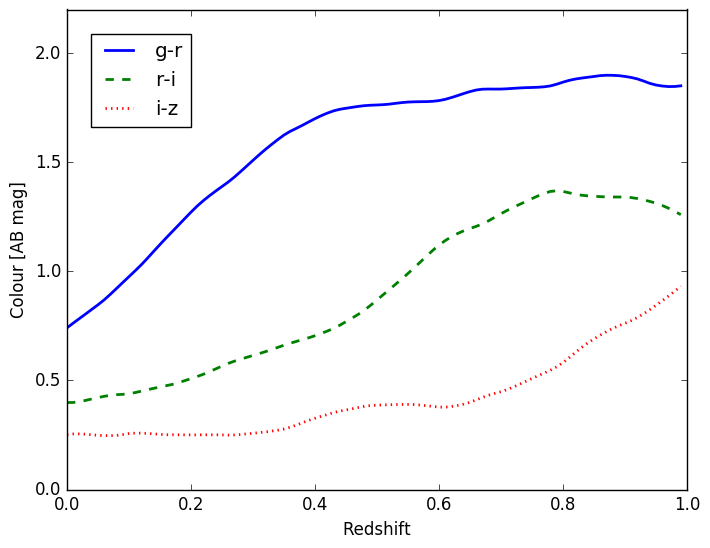}
  \caption{The expected colour-evolution of the cluster red sequence as a function of redshift for the three colours considered in the determination of photometric redshifts, i.e. $g-r$, $r-i$, $i-z$. }  
  \label{fig:sed}
\end{figure}

In order to use this technique, one needs a well calibrated model of the colour-redshift relation for the cluster red sequence. The lack of spectroscopic coverage for this sample, and the general scarcity of large, wide area spectroscopic surveys, such as SDSS (which in any case is not deep enough for our purposes), in the Southern Hemisphere means that this relation could not be derived empirically for the GROND filter set. There are however a number of publicly available spectral energy distribution (SED) templates for early-type galaxies \citep{Bruzual2003,Polletta2007,Maraston2009} which can be used to model the expected colour of the red sequence. We tested a variety of these models by comparing the templates (in the CFHTLS photometric system) with a combination of the XXL-100 brightest clusters \citep{Pacaud2016} matched with photometric redshift catalogues for individual galaxies from \citet{Mirkazemi2015}, using data from the CFHTLS wide-field surveys. We ultimately decided to use the SED of an early-type galaxy published by \citet{Polletta2007} and generated by the GRASIL code \citep{Silva1998} as this provided the best fit to the CFHTLS photometry and the lowest bias and scatter in testing on a spectroscopically confirmed subset of clusters. The colour-redshift relation for these templates was computed by making use of LePhare \citep{ilbert2006accurate,arnouts1999measuring} for the GROND filters in each of the \textit{g,r,i,z}-bands respectively. The expected colours of a typical red sequence galaxy as a function of redshift are shown in Figure (\ref{fig:sed}). 

\subsubsection{The photo-z algorithm} 
Taking advantage of information from the optical and X-ray observations, we built a `likelihood' indicator for the redshift of galaxy clusters. This function is based on the optical colour of the detected galaxies along with the position and extent resulting from the X-ray detection of the clusters. We note that this is not a true likelihood estimator but rather an empirically derived indicator for the most likely redshift of the cluster.
\begin{enumerate}
\item For each galaxy in the field-of-view, we calculate the probability that it is an early-type galaxy at a given redshift by comparing the colour of the galaxy to that expected from the SED. We assume that the scatter around the colour of the red sequence follows a Gaussian distribution with a width of 0.05 in each colour and we include the error on the photometry. The probability as a function of redshift for an individual galaxy is calculated as in Equation (\ref{gaussian}) below:

\begin{equation}\label{gaussian}
p(z) =\prod\limits_{c} \frac{1}{\sqrt{2\pi}\sigma_c}\exp\left[\frac{-(c_{gal} - c_{model})^2}{\sigma_c^2}\right],
\end{equation}
where the product runs over all colour combinations $[g-r, r-i, i-z]$, $\sigma_c = \sqrt{0.05^2+\sigma_{c,phot}^2}$, combines the width of the red sequence and the error on the photometry, $c_{gal}$ is the measured galaxy colour,  $c_{model}$ is the expected colour from the colour-redshift relations given in Figure \ref{fig:sed}.
\item This probability is then weighted by the spatial position of the galaxy relative to the X-ray centre of the cluster and the extension as calculated by the X-ray detection pipeline to give the `likelihood' that the given galaxy is a member of a cluster at that position and redshift. The selection of the X-ray centre as the cluster centre is well justified since the PSF of the XMM imaging ($\sim 15\arcsec$) is comparable to the typical size of a cluster core ($\sim$5-30\arcsec). Experimentation with various weighting schemes and beta-model exponents lead to the choice of a beta-model profile and relevant parameters given by:
\begin{equation}\label{beta}
W(r) = \left[\frac{W_0}{1+\left(\frac{r}{r_{ext}}\right)^2}\right]^{\frac{3}{2}},
\end{equation}
where $W_0$ is an arbitrary normalisation, set to unity, $r$ is the angular distance between the galaxy and the X-ray centre of the cluster and $r_{ext}$ is the angular X-ray extent, calculated from the X-ray detection pipeline.
\item This new `likelihood' is then summed over all galaxies to obtain a total `likelihood' distribution as a function of redshift for the entire cluster.
\item Additionally, the number of likely member galaxies, $N_{gal}(z)$, is calculated by selecting galaxies that have a `likelihood' indicator of more than 80\% of their peak value at each redshift and this distribution is combined with the `likelihood' indicator of the cluster to give an over all redshift distribution. 
\end{enumerate}
The final redshift `likelihood' indicator is then given by:
\begin{equation}\label{eq:likelihood}
\mathcal{L}(z) =  N_{gal}(z)\sum\limits_{gal} W(r) p(z),
\end{equation}
where $N_{gal}(z)$, $W(r)$ and $p(z)$ are as described above and the photometric redshift of the galaxy cluster is chosen such that $\mathcal{L}(z)$ is maximised. 

\subsection{Application to GROND}
\label{sub: grond-z}
\begin{figure*}
  \centering  
  \includegraphics[width=0.66\columnwidth]{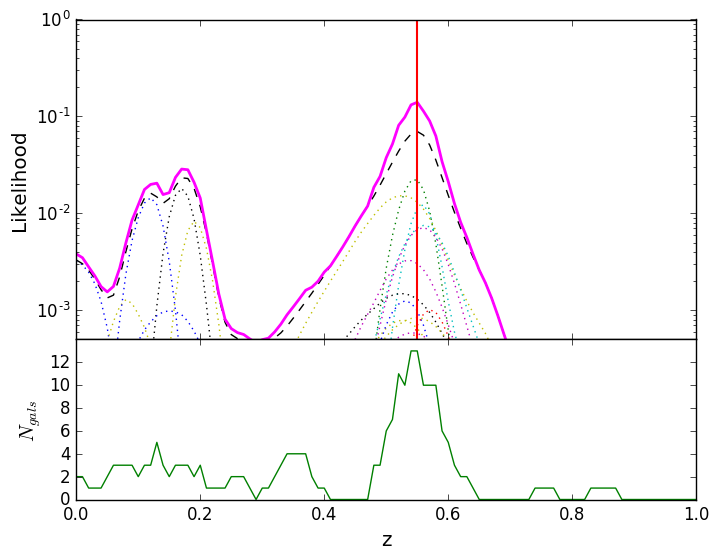}
   \includegraphics[width=0.66\columnwidth]{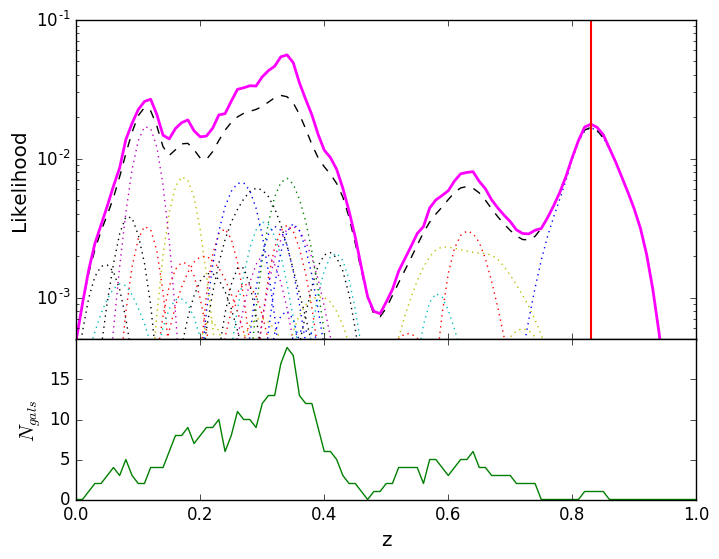}
  \includegraphics[width=0.66\columnwidth]{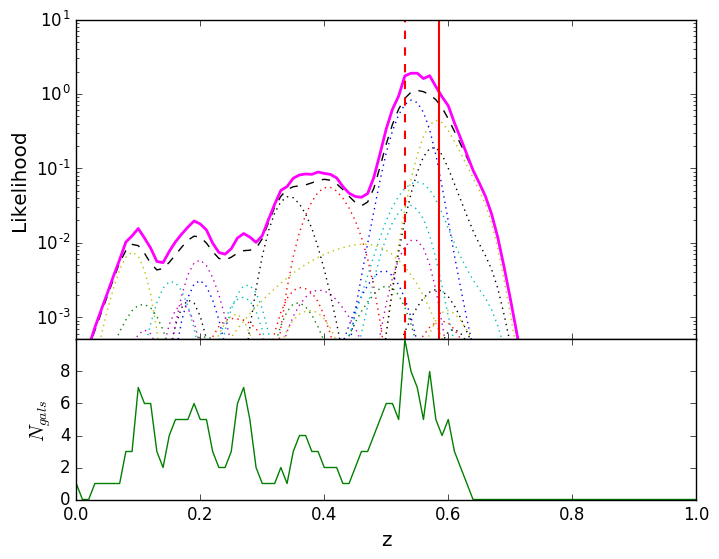}
  \caption{`Likelihood' distributions of three galaxy clusters are shown in the upper panels. \textit{Left}: X-CLASS 459, with a spectroscopic redshift $z=0.55$. \textit{Middle}: X-CLASS 228 with a spectroscopic redshift $z=0.83$ \textit{Right}: X-CLASS 430 with spectroscopic redshift $z=0.58$. The dotted lines are the `likelihood' distributions for the individual galaxies in the field, calculated from Equation \ref{eq:likelihood} and the black dashed is the $\beta$-model weighted sum of the individual galaxies. The lower panel shows the number of galaxies contributing to likelihood at each redshift. The solid magenta curve in the upper panel is the final `likelihood' given in Equation \ref{eq:likelihood}.The solid-red vertical lines indicate the redshift of the cluster obtained from the archival redshift search which, in the left and centre cases, overlap with the redshift determined after visually inspecting these curves as described in Section \ref{sub:vis_insp}. The dashed-red vertical line shows the redshift determined from the GROND observation, which is slightly different from the spectroscopic redshift of the cluster.}
  \label{fig:likelihoods}
\end{figure*}

Galaxies are selected from SExtractor source catalogues as those with CLASS\_STAR $< 0.7$, $r$-band magnitude brighter than 24.0 and signal-to-noise in the aperture defined by $MAG\_AUTO$ greater than 5.0. We use this lower value of CLASS\_STAR compared that used for the selection of stars for the astrometric calibration to reduce number of contaminating stars in our galaxy catalogues. For each of these galaxies a redshift range over which they could be possible cluster red sequence members is determined based on the criteria discussed in Section \ref{sub:contaminents}. The photometric redshift algorithm described previously is then run on each galaxy catalogue producing a `likelihood' distribution with redshift. In instances where there is more than one observation of a given cluster, a photometric redshift is calculated for each observation. The `likelihood' distributions are then compared and the best observation chosen, taking into account the `likelihood' value, the FHWM seeing of the observation and the photometric calibration of the entire night on which the observation was performed. The position of the peak value of the `likelihood' distribution is determined to be the redshift of the cluster. Examples of the `likelihood' distributions computed by our code are given in Figure \ref{fig:likelihoods} for three cases.  

\subsubsection{Removal of contaminants}
\label{sub:contaminents}
Initial testing of our method highlighted two classes of complications arising from either foreground or background contamination by galaxies not associated with cluster but along the same line of sight as the cluster centre. These contaminants are thus heavily weighted by the $\beta$-model of Equation \ref{beta}. In order to mitigate these, we defined rules to remove possible contaminating galaxies which would otherwise strongly, and negatively affect our redshift calculations. These constraints were then used to pre-filter that galaxy catalogues before entering the photometric redshift algorithm. 

The first class of impediments was the presence of distant, star-forming galaxies with similar apparent colours to a lower redshift early-type galaxy. To remove these, we selected galaxies based on the $r$-band magnitude-redshift relation. The magnitude, $m_*(z)$, was computed as a function of redshift using a \citet{Bruzual2003} stellar population model. This model was fixed to a single burst of star formation at $z=3$, with solar metallicity and Salpeter initial mass function \citep{Salpeter1955} , and evolved through redshift space by making use of the publicly available \texttt{EzGal} package \citep{Mancone2012}. Following the methodology of \citet{Rykoff2012} and \citet{Mirkazemi2015}, $m_*(z)$ was normalised such that $m_{*,i'}(z=0.2)=17.85$ in the SDSS filter system, corresponding to a galaxy with luminosity $L_*=2.25\times10^{10}L_{\odot}$. Thus, any galaxy fainter than $m_{*,r}(z)+2.5$ was excluded from the likelihood calculation. 

The second class was due to galaxies that had a single colour agreeing well with that expected from the SED of an early-type galaxy while the other two colour constraints were only marginally met, implying that these were unlikely to actually be cluster red sequence members. These galaxies were eliminated by placing constraints on the colour allowed for the individual galaxies in multiple bands. In order to have sensitivity to the $4000\AA$  break over a wide range of redshifts and to enhance the robustness of the selection, possible member galaxies were constrained to be those with $g-i$ and $r-z$ colours consistent with those described in the previous section. This step was meant to eliminate only obvious contaminants and as such a broad range of allowed colours was chosen, so that only galaxies with a colour within 0.5 of that expected from the model were included in the redshift calculation. 

\subsubsection{Visual inspection of results}
\label{sub:vis_insp}
Since the number of clusters to be followed up is relatively small, and we are working with pointed observations, it is possible to visually inspect every cluster candidate. Once every cluster had a single redshift assigned to it, a visual inspection by three astronomers (J. Ridl, N. Clerc and J. Sanner) was performed. The results from running the photometric redshift algorithm, (see examples in Figure \ref{fig:likelihoods}) are compared with three-colour ($gri$) images, and images in which the most likely redshift for individual galaxies, assuming them to be early-type galaxies, is over-plotted. We are thus able to check that the output photometric redshift of the photo-z algorithm matches what would be roughly expected by a human eye and obvious errors can be corrected. This happens most frequently for high-redshift clusters, where the number of cluster members detected is very low. It is thus far easier for the result to be contaminated by a foreground elliptical galaxy nearby in projection to the X-ray center. Additionally, some measurements were affected by a very bright, saturated star or a secondary reflection from a nearby bright star, close to the X-ray centre of the cluster which causes a large fraction of the cluster members to be excluded from the calculation. 

This visual inspection procedure found that in 37 out of 266 cases the photometric redshift pipeline had selected an incorrect peak in the likelihood distribution, typically due to contamination by a foreground galaxy resulting in a significantly lower redshift being reported than that expected from the visual appearance of the apparent cluster members and their distribution. For these cases, the position of the peak was remeasured after removing the contaminating source. We also identify a subset of 19 clusters as being likely distant $z>0.8$ candidates, which we discuss in Section \ref{sub:distant}. Any prior knowledge of the redshift of the clusters from the archival matching was hidden from the inspectors which is important for validating the visual inspection process over the entire sample. 

The examples presented in Figure \ref{fig:likelihoods} illustrate three typical cases. For the first cluster, X-CLASS 459, there is a clearly defined peak which all three inspectors agreed was correct. It turns out to match the spectroscopic redshift of $z=0.55$ \citep{Barcons2007} in the literature to within $\delta z = 0.01$. The second example, X-CLASS 228, is one where all inspectors agreed that the most likely redshift of the cluster lies around the peak at $z\sim 0.8$. Initially, the photometric redshift algorithm determined the redshift to be $z=0.34$. The visual inspection however revealed that this measurement was likely affected by the presence of a foreground cluster (X-CLASS 229) at a distance of 2 \arcmin\ away. Visually, the mostly likely peak from the likelihood indicator appeared to be the one around $z \sim 0.8$ and the redshift was re-measured around this peak resulting in a redshift of $z=0.83$, in agreement with the  redshift provided by the XMM Distant Cluster Project \citep[XDCP][]{Nastasi2014}. The final example, X-CLASS 430, is a difficult case as two peaks appear nearby to one another in the likelihood distribution. In such cases we decide to trust the maximum likelihood peak as being the redshift of the cluster at $z=0.53$ but for this example, when comparing to the spectroscopic redshift $z=0.585$ \citet{Guennou2014structure}, we find that the redshift has been underestimated and the higher peak should have been selected.  

\subsection{Unconfirmed clusters}
\label{sub:unconfirmed}
Apart from the clusters identified as being distant candidates, we are further unable to confirm the redshift for 10 clusters for a variety of reasons. It was found to be impossible to observe X-CLASS 51 due to the presence of a very bright star in the GROND field-of-view. We were also unable to obtain an observation of sufficient quality for X-CLASS 2311 due to the lack of a usable guide star on which GROND could track. The X-ray detection of X-CLASS 560 is heavily contaminated by an AGN and no obvious red sequence of galaxies is seen in the GROND observation. We were unable to reach consensus as to whether or not this is a distant candidate. We were unable to obtain a redshift for X-CLASS 1400 as the only available observation took place on a night with an insufficiently good photometric calibration. We do however see a clear red sequence of galaxies and estimate the redshift visually to be $z\sim 0.7$. X-CLASS 1995 and 2002 are both affected by the presence of bright stars which prevent the recovery of the photometric redshift. For X-CLASS 996, 997, 998 and 2078 we are unable to obtain a suitable astrometric solution due to the lack of enough viable stars in the optical field-of-view of the observations. 

\subsection{Comparison between GROND and archival redshifts} 
\label{sub: z-comp}
\begin{figure}
  \centering  
  \includegraphics[width=0.99\columnwidth]{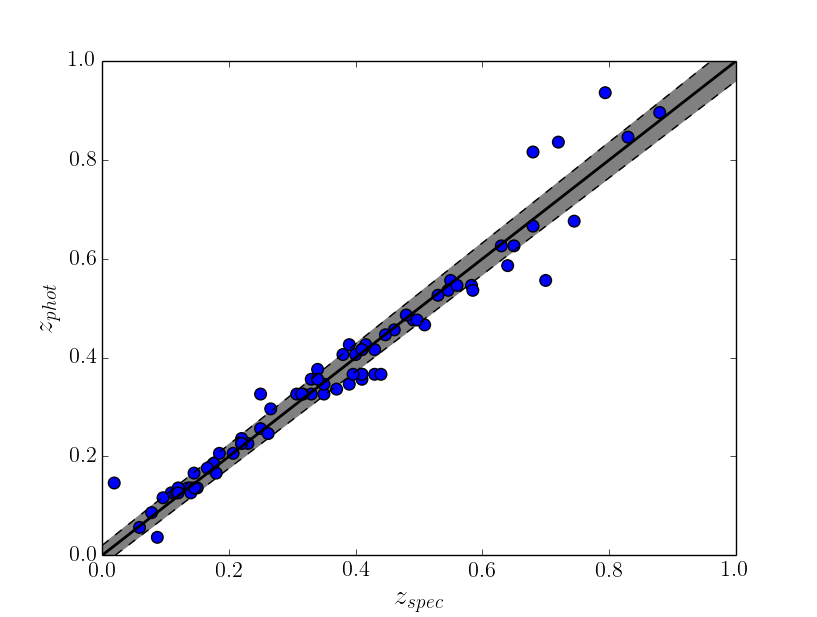}
  \caption{Comparision of GROND photometric redshift with 76 of the spectroscopic redshifts with $z < 0.85$ obtained from the literature as discussed in Section \ref{sub:archival}. The grey shaded region is bounded by the lines $ z \pm 0.02(1+z)$, indicating the typical error of our measurements. }
  \label{fig:z_comp}
\end{figure}

In order to validate our photometric redshifts, we compare them with the sample of 76 spectroscopically confirmed galaxy clusters from various sources, as discussed in Section \ref{sub:archival}. We notice that the scatter around the one-to-one line in Figure (\ref{fig:z_comp}) increases around a redshift of $z \sim 0.4$. This is due to the fact that the 4000 \AA\ break moves from the $g$ to $r$-band filter, increasing the uncertainty in the colour-redshift relation at this point. We also note that our method is unable to compute reliable uncertainties for the photometric redshift determined from Equation \ref{eq:likelihood} and so we do not provide errors for individual cluster measurements. We are only able to give an indication of the average error for the entire sample. We find that our redshifts are accurate to $\Delta z = 0.02(1+z)$. Practically all of the constraining power of z-CR-HR method, for which this sample has been assembled, is provided by binning clusters in redshift bins of $\Delta z = 0.1$ \citep{Clerc2012a}. Our redshifts are thus of a suitable quality in order to proceed with a cosmological analysis (Ridl et al., in prep). 

\section{Characterisation of X-ray properties}
\label{sec:x-ray}
\subsection{Growth curve analysis}
\label{sub:gc}
The first step in determining the X-ray properties of detected galaxy clusters is to measure the X-ray detector count-rate. For the sample presented in this paper, count-rate measurements had already been performed in multiple bands as a function of radial distance from the X-ray defined centre of the cluster. A semi-iterative method is used to deal with sources that either occupy a large fraction of the detector or are heavily contaminated by point-souces such as AGN and allows for the manual redefinition of the cluster centre.

Count-rates, defined as the mean number of photons detected by the CCDs in one second, are measured in concentric annuli under the assumption that the source is spherically symmetric. This provides a straightforward way to correct for masked point sources, CCD gaps or detector borders, where part of the cluster lies outside the field-of-view of one of the cameras. These are then corrected for vignetting, and are thus equivalent to having the source positioned at the centre of the camera. The count-rates are always calculated on the full exposure of the given pointing, as opposed to the 10ks or 20ks subsets used for the source detection, ensuring a maximal signal-to-noise for each measurement. Each of the detectors is treated independently and the individual count-rates summed giving a total growth curve as a function of radius.  

These measurements were validated through the use of simulated XMM observations of clusters and all count-rates were corrected for the fact that XMM observations are performed with the use of different filter (THIN1, MEDIUM or THICK) configurations at the discretion of the guest observer. For further details, see Section 2.4 of \citet{Clerc2012}.

\subsection{Energy conversion factors}
\label{sub:ecf}
In order to convert the observable, count rate into flux it is necessary to determine an energy conversion factor (ECF). This was accomplished by selecting a set of 8 XMM observations spanning the 2000-2010 period, in order to test the long term variation of the ECF. These were used as representations of prototypical X-CLASS pointings. Since all count rates are equivalent to being on-axis cluster observations, we calculate the ECF for each respective observation only at the centre of each of the MOS and PN cameras.

A key step in the calculation of the ECF for a given observation is to create the ancillary response file (ARF) and redistribution matrix file (RMF). The observations were downloaded from the XMM Science archive\footnote{http://nxsa.esac.esa.int/nxsa-web/} and the standard preliminary data reduction performed as detailed in the XMM data analysis manual including running the SAS tools \texttt{cifbuild} and \texttt{odfingest}, making use of the XMM calibration repository locally available at MPE. The data were then processed for the MOS1, MOS2 and PN chips individually, by running the tools \texttt{emproc} and \texttt{epproc} respectively to produce calibrated event lists. Light curves were then extracted and used to create good-time-intervals (GTIs) and these were used to remove periods heavily affected by proton and solar flares from the calibrated event lists. Finally, the SAS tools \texttt{rmfgen} and \texttt{arfgen} were used to create the RMF and ARF respectively. 

Next, we used PyXspec \citep{arnaud1996xspec} to compute the energy conversion factors by simulating XMM observations of model galaxy clusters with a range of temperatures from $T = [0.3-10]$ keV, hydrogen column densities from $\mbox{nH} = [0.01-0.2] \times 10^{22}cm^{-2}$ and redshifts from $z = [0.05-1.2]$. For each iteration, an observation was simulated using the PyXspec function \texttt{fakeit} making use of the RMF and ARF described above on each of the cameras individually, and using an exposure time of $10^7$s to limit the Poisson errors inherent in X-ray observations. The normalisation of the PyXspec model was chosen such that a cluster with $T=1.0$ keV, metallicity $Z=0.3$ Z$_{\odot}$ and redshift $z=0.1$ would have a flux of $10^{-13}$ erg $\mathrm{s^{-1}cm^{-2}}$. We then selected the channels corresponding to the energy range of interest, i.e. [0.5-2] keV, and computed the count rate in this energy band. This count rate was then compared with the model flux to give the necessary multiplicative factor to convert between the two quantities for each camera independently. These individual factors were then inverse summed giving the energy conversion factors on a grid of temperatures, hydrogen column densities and redshifts. 
 
\subsection{Physical parameter measurements}
\label{sub:xray_measurements}
The physical parameters such as X-ray luminosities, temperatures, cluster masses and the radius at which the average density of a cluster is 500 times the critical density of the Universe, $r_{500}$ are calculated using an iterative method, similar to that of \citet{vsuhada2012xmm}.  This method is summarised below with initial values of $T_{300 kpc} = 2.5$ keV and $r_{500}=0.5$ Mpc respectively.
\begin{enumerate}
\item The value $r_{500}$ is converted from Mpc into arcseconds making use of the Astropy Cosmology module, which allows for straightforward cosmological calculations.
\item The count rate enclosed by this radius is extracted from growth curves, as presented Section \ref{sub:gc}. 
\item We next convert this count rate to X-ray flux, making use of the relevant energy conversion factor as described in Section \ref{sub:ecf} depending on the cluster redshift, the hydrogen column density of the pointing and the current value of the temperature.
\item The X-ray luminosity $L_{500}^{[0.5-2]keV}$, in the [0.5-2.0 keV] band is then calculated along with the bolometric ([0.05-100] keV) luminosity by making use of PyXspec, the Python implementation of XSPEC. To do this, we assume an absorbed APEC (\texttt{phabs*apec}) model with the following model parameters: hydrogen column density set to the value calculated at the position of the pointing; temperature set to the current $T_{300 kpc}$ value; metallic abundance 0.3Z$_{\odot}$, redshift set to the spectroscopic redshift where available (i.e. redshift type: confirmed) or the photometric redshift calculated from the GROND observations as described in Section \ref{sec:redshifts}. The normalisation is set such that the flux in the [0.5-2] keV band matches that calculated in step (iii) above. The function \texttt{calcLumin} is then used to determine the cluster luminosity in the [0.5-2] keV and [0.05-100] keV bands.
\item The scaling relations derived by the XXL \citep{Pacaud2016,Giles2016,Lieu2016} are utilised to obtain the temperature within 300 kpc ($T_{300 kpc}$) and $M_{500}$,\footnote{$E(z)^2 = \Omega_M(1+z)^3 + \Omega_\Lambda$}
\begin{align}
\frac{L_{500}^{[0.5-2]keV}}{3\times10^{43} \mbox{erg s$^{-1}$}} = 0.71 \left(\frac{T_{300 kpc}}{\mbox{3 keV}}\right)^{2.63} E(z)^{1.64},\label{eq:L-T}\\
\frac{M_{500}}{2\times10^{14}\mbox{M$_{\odot}$}} = 1.16 \left(\frac{T_{300 kpc}}{\mbox{3 keV}}\right)^{1.67} E(z)^{-1}.\label{eq:M-T}
\end{align}
\item Finally, a new value for $r_{500}$ is calculated from the relation, \footnote{$\rho_c = E(z)^2 3 H_0^2 / 8\pi G$}
\begin{equation}
M_{500} = 500 \rho_{c}\times\frac{4\pi}{3}r_{500}^3.
\end{equation}
\item Steps (i)-(vi) are the repeated with the updated values for $T_{300 kpc}$ and $r_{500}$ until the calculated value for the temperature has converged to an accuracy of 0.01 keV. 
\end{enumerate}  
For 3\% of clusters with a reliable redshift, this method does not converge. These failures are either distant ($z > 1$) clusters or very nearby and contaminated by X-ray emission from the BCG, as discussed in Section \ref{sub:groups}.

\subsection{Errors on X-ray derived properties}
\label{sub:errors}
For the values calculated for the X-ray parameters in this paper, we consider only errors introduced by the uncertainty in the measured count-rate in the [0.5-2] keV band, the error in the redshift assigned to the cluster and the scatter around the $L-T$ and $M-T$ scaling relations. We determine the uncertainly introduced by each of these parameters by offsetting their values, one-by-one, by $1\sigma$ for the count-rate and scaling relations and by the average error, $\Delta z = 0.02 (1+z)$, for the redshift in the iterative process described in the previous section. The uncertainties for all quantities calculated in the iterative process e.g. $L_{500}^{[0.5-2]\mathrm{keV}}$, but here we discuss only the errors on the bolometric luminosity and temperature since these are the quantities which we compare with already existing measurements provided by the XMM-XXL and XMM-XCS catalogues.  

We find that the dominant source of uncertainty in the calculated properties comes from the scatter on the $L-T$ relation, where we find that on average the calculated value for the bolometric luminosity is offset by $\sim 20\%$ and the temperature by $\sim 33\%$. The other parameters all influence the measurements by less than 10\% apart from the redshift uncertainty which introduced an error of $\sim 14\%$. The final error bars shown in all plots containing the X-ray properties calculated in this work are determined by summing the individual errors in quadrature. The results of the error calculations are summarised in Table \ref{tab:xray_errors}. 
\begin{table}
\begin{center}
 \caption{Average errors induced by offsetting the X-ray count-rate and redshift of the clusters and adjusting the scaling relations by their respective scatter and their effect on the bolometric luminosity and temperature obtained from the iterative method. The totals are calculated by adding the individual errors in quadrature.}
    \begin{tabular}{|c|c|c|}   
   
    	Parameter & $ \sigma_L$ & $ \sigma_T$ \\
	  \hline
	$\sigma_{L-T}$&  20\% &  33\%  \\
	 $\sigma_{M-T}$ & 6\% & 2\%\\
	 Count-rate &  9\% &  3\%  \\
	 $\Delta z$ & 14\% & 3\%\\
	 \hline
	 Total & 27\% & 34 \% \\
	 \hline
	 \label{tab:xray_errors}
    \end{tabular}
    \end{center}
\end{table}

\section{Results}
\label{sec:results}
\subsection{Spatial distribution of clusters}
\begin{figure*}
  \centering  
  \includegraphics[width=1.99\columnwidth]{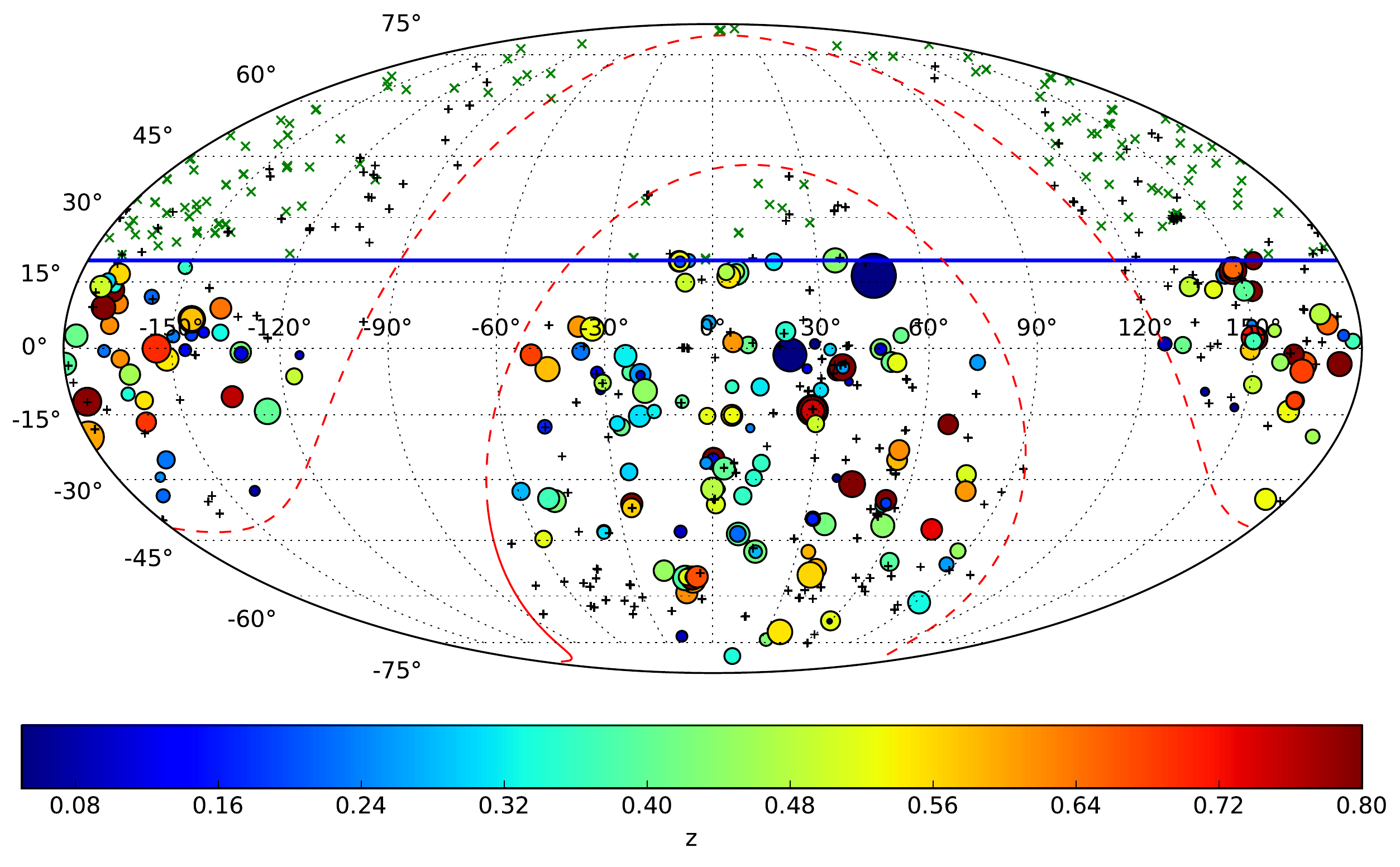}
  \caption{Distribution of clusters across the sky. X-CLASS clusters presented in this sample are described by coloured circles. The colour of the marker indicates the redshift on the cluster and size is proportional to the X-ray luminosity. The X-CLASS clusters further north than the limits of this survey are indicated by green x's and candidate clusters from a new processing of XMM data up to January 2012 are indicated by black $+$'s. The solid blue line shows the declination above which we do not observe and the red curves show Galactic latitudes $b = \pm 20\degr$. Coordinates are given in the Equatorial J2000 system. }
  \label{fig:sky_distribution}
\end{figure*}

The selected XMM pointings are distributed throughout the high-Galactic latitude sky as illustrated by Figure \ref{fig:sky_distribution}. As such, cluster number densities and distributions in various parameter spaces should be only minimally affected by cosmic variance. In principle the density of detected clusters on the sky should continue to increase with future iterations of X-CLASS, for as long as XMM continues to function normally.  Already a processing of new pointings up to January 2012 (Faccioli et al., in prep) has added an additional $\sim 184$ cluster candidates (72 or which already have redshifts), shown on Figure \ref{fig:sky_distribution}. So long as systematic followup of these new clusters is available, X-CLASS will remain a competitive cosmological sample for the near future and provide an excellent compliment to future surveys with eROSITA onboard SRG. 

\subsection{Redshift distribution of clusters}
\label{sub:redshift_distribution}
\begin{figure}
  \centering  
  \includegraphics[width=0.99\columnwidth]{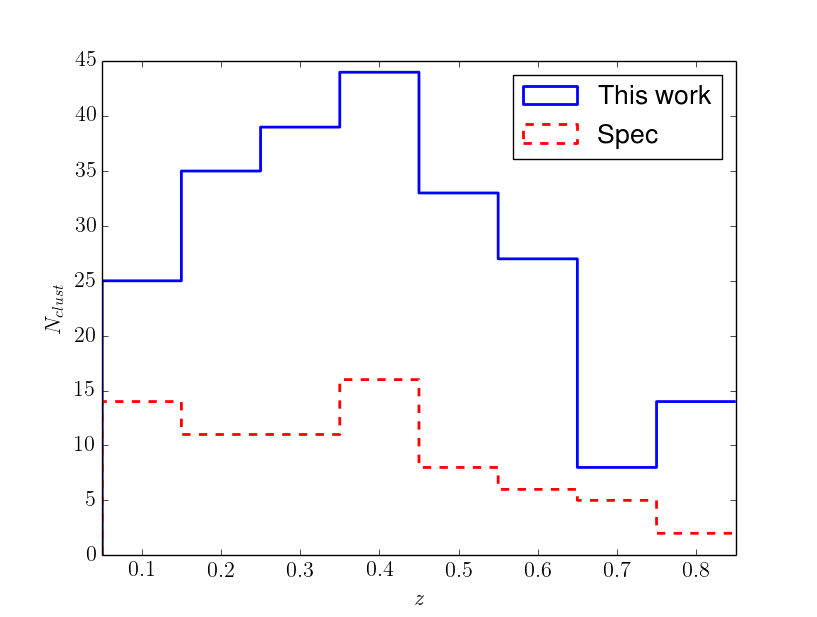}
  \caption{Distribution of X-CLASS clusters redshifts: GROND photometric redshifts for all clusters in the southern X-CLASS cosmological sample (solid-blue) and the spectroscopically confirmed subsample (dashed-red).}
  \label{fig:z_hist}
\end{figure}

As stated earlier, the number density of clusters as a function of redshift depends strongly on the underlying cosmological model. The distribution of clusters with redshift as computed in this sample is displayed in Figure \ref{fig:z_hist}. For comparison, we also plot the distribution of clusters classified as `confirmed' (spectroscopic) in the comparison with archival redshifts. We find good agreement between the two sets of redshifts.  

We find that the median redshift for the X-CLASS sample is $z=0.37$ when assigning a lower limit of 0.85 to all clusters which were classified as being `too distant' to obtain a redshift with a single 20 min OB, compared with $z=0.33$ for XXL-100 and $z=0.30$ for XCS-DR1. The difference with XXL-100 probably arises from the fact that their sample is based on a significantly higher flux limit than that inherent in our sample and thus a smaller fraction of distant clusters are included in their sample. The XCS-DR1 on the other hand, includes more XMM pointings, including those not included in this analysis due to insufficient exposure times. As such they detect more small, low-redshift  groups, thus increasing their fraction of low redshift clusters. 

The typical error on the redshift is found to be $\Delta z =0.02(1+z)$ and the outlier fraction, defined as having $|z_{grond}-z_{spec}|>3\Delta z$ is 5\%. In addition to providing redshift for 244 clusters of galaxies, we were able to provide lower limits on the redshifts of 18 distant cluster candidates. We consider a cluster to be ``too distant" since the X-ray emission is clearly extended, by virtue of the C1$^{+}$ classification but we do not find any appreciable red sequence consistent with it. It is possible that these cluster candidates are spurious detections and only the inspection of deep optical/IR imaging and/or deep Chandra observations could confirm the true nature of these objects. We discuss this further in Section \ref{sub:distant}. We also find 10 clusters with a redshift $z\la 0.1$. These clusters represent an interesting subsample as it is difficult to measure their X-ray properties and we enter a more detailed discussion of this in Section \ref{sub:groups}. 

\subsection{X-ray properties of X-CLASS}
\label{sub: l-z}
\begin{figure*}
  \centering  
  \includegraphics[width=0.99\columnwidth]{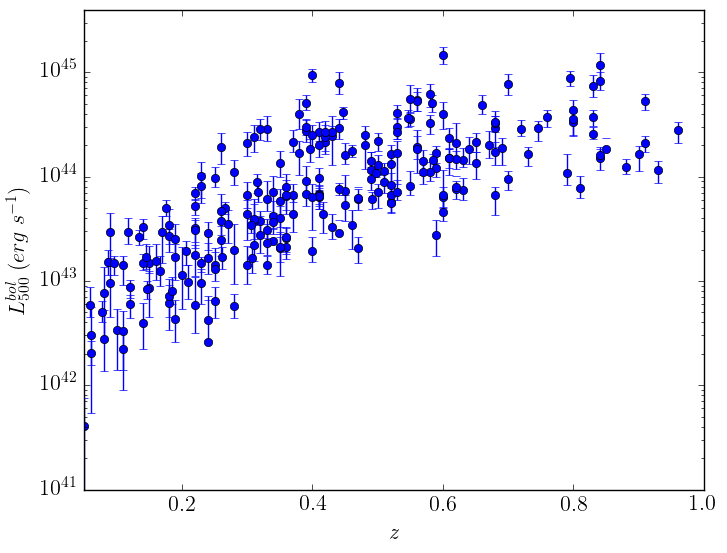}
  \includegraphics[width=0.99\columnwidth]{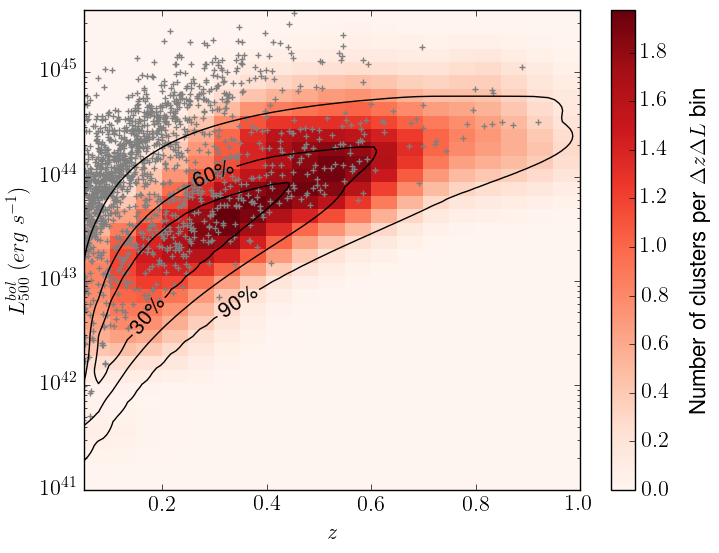}
  \caption{\textit{Left}: The distribution of X-ray luminosity as a function of redshift for X-CLASS clusters. \textit{Right}: The number density of X-CLASS clusters as a function of X-ray luminosities with redshift are indicated by the colour map, smoothed with a Gaussian filter. The contours indicate the expected distribution from the eROSITA 4 year all-sky survey under the assumptions discussed in the text and the grey $+$'s represent the ROSAT selected MCXC meta-catalogue \citep{piffaretti2011mcxc}. }
  \label{fig:z_Lx}
\end{figure*}

An important characterisation of a sample of X-ray selected galaxy clusters is the relationship between the cluster redshifts and their luminosities as it gives an indication of the mass range represented by the sample. The distribution for this sample is shown in Figure \ref{fig:z_Lx}. We also plot the expected cluster distribution from the full eROSITA all-sky survey (eRASS), with a selection function based on realistic eRASS simulations \citep{Ramos2016}, and using the XXL scaling relations \citep{Pacaud2016,Giles2016,Lieu2016}, WMAP9 cosmology \citep{hinshaw2013wmap} and the Tinker mass function \citep{tinker2008toward}. For reference we also show the distribution of the MCXC cluster sample which is based on the ROSAT All-sky survey and serendipitous cluster catalogues \citep{piffaretti2011mcxc}.  We notice that we detect fewer high luminosity clusters at low redshifts. The reason for this is two-fold. Firstly, the number of luminous clusters is limited at low redshifts due to the smaller volume which is probed compared to higher redshifts, and secondly, because very massive, nearby clusters have been deliberately excluded from the sample. From the right panel of Figure \ref{fig:z_Lx} we see that on average X-CLASS probes slightly higher redshifts and X-ray luminosities than expected from eROSITA. 

\begin{figure*}
  \centering  
  \includegraphics[width=1.99\columnwidth]{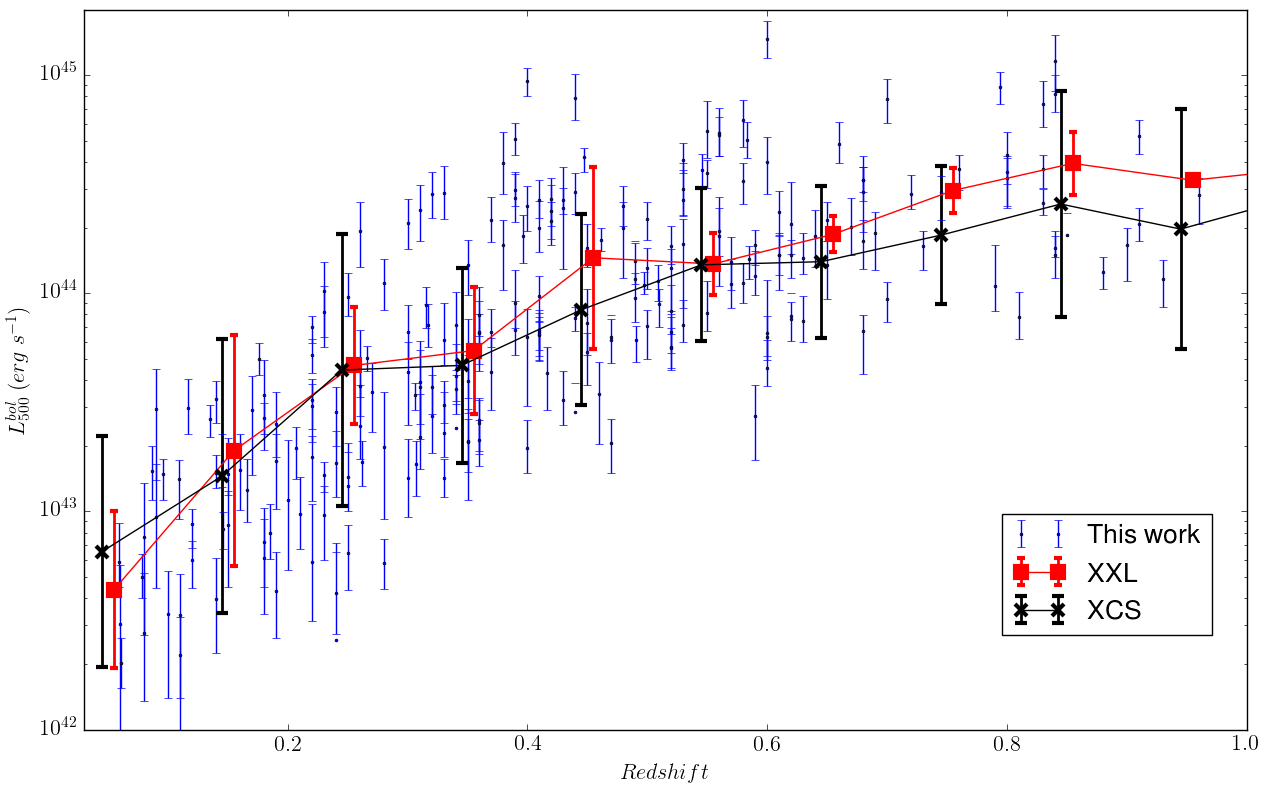}
  \caption{Distribution of X-ray luminosities as a function of redshift for X-CLASS clusters compared to the XCS-DR1 and XXL-100 catalogues. The XCS and XXL catalogues have been binned by to redshift slices of width $z=0.1$ and the error bars represent the respective scatter about the median luminosity of each bin.}
  \label{fig:z_Lx_2}
\end{figure*}

\begin{figure}
\centering
\includegraphics[width=0.99\columnwidth]{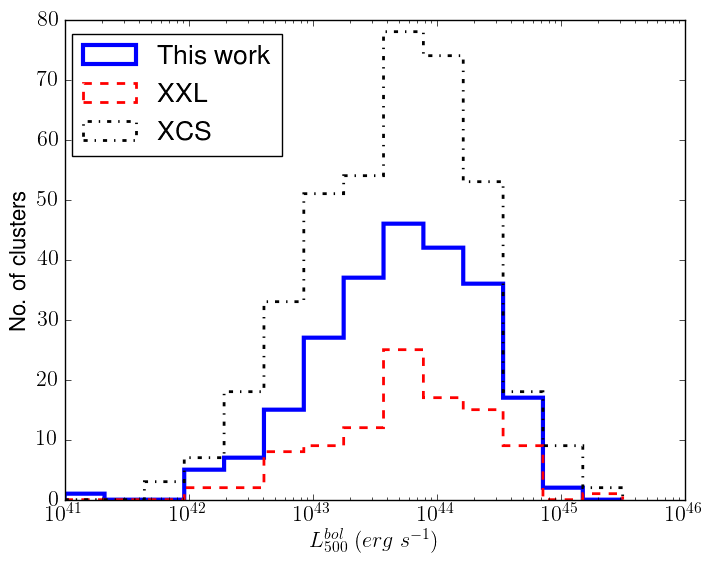}
\caption{The number of clusters as a function of bolometric luminosity for the X-CLASS sample presented in this paper (solid blue line), compared with the distributions of the XCS-DR1 (black, dashed-dot) and XXL-100 (red, dashed) samples respectively.}
\label{fig:lum_hist}
\end{figure}

It is also useful to see how this sample compares with other similar XMM surveys. In Figure \ref{fig:z_Lx_2} we show the X-CLASS luminosities as function of redshifts along with those from the XXL-100 and XCS-DR1 catalogues overlaid. The distribution of the X-ray bolometric luminosity of these three samples is displayed in Figure \ref{fig:lum_hist}. These two plots illustrate some interesting differences between the samples. We notice the high number of bright nearby objects relative to our sample as expected from our removal of sources with high ($> 0.5\ \mathrm{cts\ s^{-1}}$) count rates. The lower flux limit of the XCS-DR1 is also clearly apparent. As expected we probe a significantly lower luminosity range than the XXL-100 although we would expect a more similar lower flux limit when compared to the entire XXL-C1 cluster sample consisting of 267 spectroscopically confirmed clusters which is yet to be released (Adami et al., in prep). The deficit in the number of high luminosity, high redshift clusters in the X-CLASS sample compared (in particular) to the XXL-100 is largely due to the fact that we do not have a secure redshift for many clusters with $z > 0.85$ and have relied on either photometric, or where available, spectroscopic redshifts already existing in the literature.  

Ultimately, X-CLASS seems to be complementary to the XXL-100 and XCS-DR1 samples. Although not pushing to fluxes as low as the XCS-DR1, the decision to fix the exposure times to 10ks or 20ks greatly simplifies the selection function. Given that the (almost)-identical detection algorithm is used for the XXL and X-CLASS, we expect that the final XXL-C1 sample should have similar properties to the one presented here. While the XXL will not be affected by biases arising from including pointed observations of already known clusters, X-CLASS is assumed to be less affected by cosmic variance due to its scattered nature across the sky and has the potential to probe a significantly larger area of the sky. Much of the area covered by X-CLASS however lacks overlap with homogeneous and deep multi-wavelength surveys and followup, which this paper partially addresses. 

\subsection{The X-CLASS/GROND cluster catalogue}
We present the X-ray selected, X-CLASS/GROND cosmological catalogue in Table \ref{tab:catalogue} in Appendix B. Column 1 in Table \ref{tab:catalogue} is the X-CLASS catalogue ID. Columns 2 and 3 give the right ascension and declination of the X-ray centroid respectively. The photometric redshift, as derived from GROND observations is provided in Column 4. Where available, Columns 5 and 6 contain the redshift of the cluster as recovered by cross-matching the X-CLASS catalogue with various catalogues, such as XCS-DR1, redMaPPer and others in the NED, and the status flag of this redshift, as described in Section \ref{sub:archival}. Column 7 contains the count rate, given in units of counts per second, of the cluster in the [0.5-2 keV] band.  Columns 8-10 contain various physical properties of the clusters calculated in Section \ref{sec:x-ray}, namely $r_{500}$,  $L_{500}^{[0.5-2] keV}$, the luminosity in the [0.5-2 keV] band, measured in units of $10^{43}\ \mathrm{erg\ s^{-1}}$ in an aperture out to $r_{500}$; and finally the temperature of the cluster derived from the XXL scaling relations (Equations \ref{eq:L-T} and \ref{eq:M-T}) in keV.


\section{Discussion}
\label{sec:discussion}
\subsection{Comparison of X-ray measurements with other XMM surveys}
In order to ensure that we were able to accurately recover the X-ray properties of our sample, we compared the results of the analysis presented in Section \ref{sec:x-ray} to the results obtained by the XXL and XCS teams. Since, the XCS-DR1 catalogue contains only bolometric luminosities we compare these, as opposed to luminosities in the [0.5-2] keV rest-frame luminosities. Due to the fact that our cluster temperatures are calculated from the $L-T$ scaling relation given by \citet{Pacaud2016}, we expect that the quality of the fits of luminosity and temperature should be strongly correlated in the comparison with the XXL-100, i.e. a good agreement between the luminosities should provide good agreement between the temperatures. An important difference between the calculations presented here and those of the XXL-100/XCS-DR1 samples is that in the latter analyses, X-ray physical parameters were calculated through spectral template fitting directly to the X-ray data as opposed to the iterative method presented in Section \ref{sub:xray_measurements}. Spectral template fitting is considered to be the ``gold standard" method for obtaining cluster temperature and luminosities and this forms the basis of a currently ongoing study (Molham Mostafa et al., in prep). For the purposes of this paper, we deemed it sufficient to use the much faster iterative method, which as shown by \citet{vsuhada2012xmm} gives suitably accurate results and allows for a good characterisation of the overall sample. 

The matching between X-CLASS and XXL-100/XCS-DR1 was done through the use of \texttt{TOPCAT} with a matching radius of 2\arcmin. This radius was chosen because it was found to be large enough that it is able to account for the differing definition of the cluster centres given in the catalogue arising from the slightly different detection and measurement algorithms, and small enough that unrelated clusters were not matched to one another by chance. We found 11 and 64 clusters in common with the XXL-100 and XCS-DR1 catalogues covering a range of luminosities from $8\times 10^{42} - 5\times10^{44} \mathrm{erg\ s^{-1}}$ and $2\times 10^{42} - 10^{45} \mathrm{erg\ s^{-1}}$ respectively. 

\begin{figure*}
  \centering  
  \includegraphics[width=0.99\columnwidth]{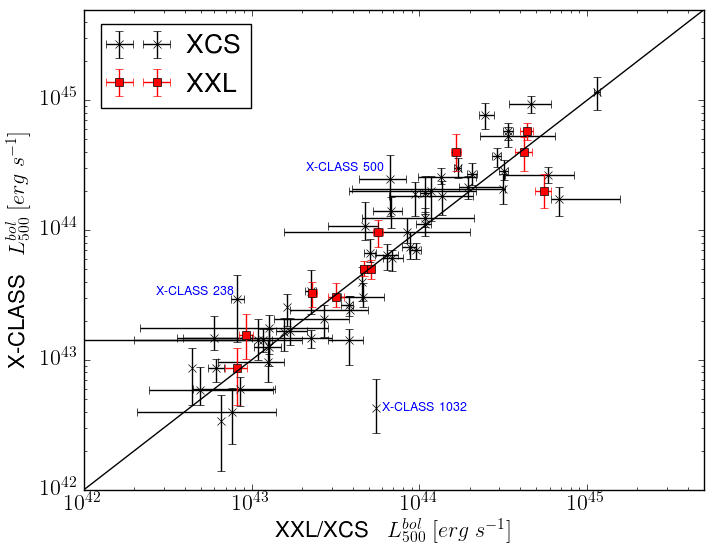}
  \includegraphics[width=0.99\columnwidth]{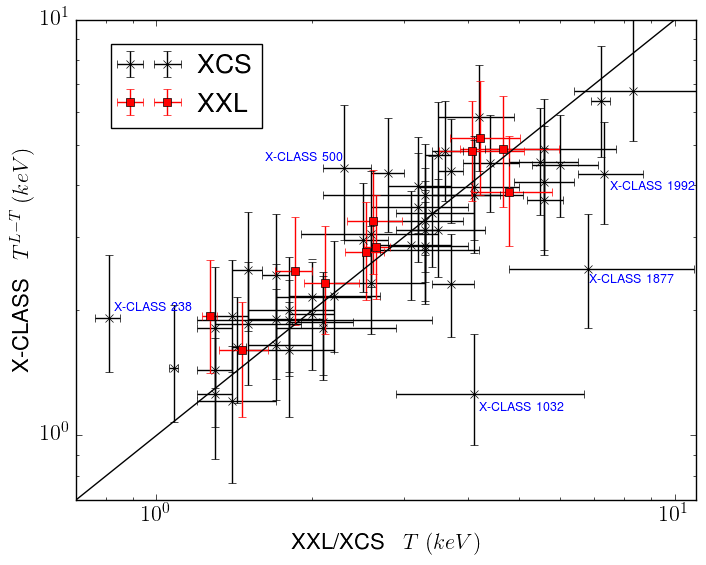}
  \caption{Comparision of X-CLASS bolometric [0.05-100 keV] X-ray luminosities within $r_{500}$ of the cluster centre (left) and the X-ray temperatures(right) with the same quantities from the XCS and XXL catalogues.}
  \label{fig:lum_comp}
\end{figure*}

Figure \ref{fig:lum_comp} shows the good agreement between the values calculated for the bolometric luminosity and temperature respectively. The bias and standard deviation of the fit between the X-CLASS and XXL-100/XCS-DR1 calculated physical properties are summarised in Table \ref{tab:fit_xray}. The good agreement with the XXL catalogue is somewhat unexpected given the similar nature of the processing, and that the luminosity and temperature measurements presented here are based on the XXL-100 scaling relations. The comparison with XCS-DR1 is a more reliable test of our measurements as they are computed by a completely independent team with different detection and measurement tools. We notice that the scatter around the one-to-one line is greater when comparing to XCS-DR1 that when compared to XXL-100. This is to be expected given that the XXL-100 measurements are performed on a significantly higher signal-to-noise sample, reflected in the size of the error bars. 
 
The comparison with bolometric luminosities and more noticeably the temperature with XCS-DR1 highlight a number of clusters for which measurements are difficult for a variety of reasons. We performed further calculations based spectral fitting to resolve the tensions between the temperatures calculated in our analysis and those presented by XCS. We find that for X-CLASS 1032 (XMMXCS J0959.5+0526) the temperature recovered from our spectral analysis are in tension with those of XCS. For X-CLASS 1992 (XMMXCS J0959.6+0231) we find that our measurement is strongly affected by a high off-axis position on a pointing with a 20 ks exposure whereas the XCS measurement is performed on a pointing with the source more centralised but only 10 ks exposure. X-CLASS 1877 (XMMXCS J1000.4+0241) appears to be a rather complicated system and is likely affected by projection effects. It is originally detected at a similar redshift to the one we calculate here ($z=0.35$) in an XMM survey of the COSMOS field \citep{Finoguenov2007}. Subsequently, numerous large-scale structures have been reported within 1\arcmin\ at redshift $z\sim0.7$ \citep{wen2011galaxy,sochting2012ultra} and so it is likely to be difficult to accurately measure the X-ray emission associated with the cluster at $z=0.35$. The measurement of X-CLASS 238 (XMMXCS J0000.4-2512) is probably affected in our case by additional counts entering the calculations due to the presence of a nearby Abell cluster, A2690, which was the original target of the observation. Finally X-CLASS 500 (XMMXCS J0306.2-0005) is probably affected by a relatively high background in the pointing and nearby point sources.

\begin{table}
\begin{center}
 \caption{The bias and standard deviation of a comparison with other XMM cluster surveys.}
    \begin{tabular}{|c|c|c|c|}   
   
    	Catalogue &\ & $L_{500}^{bol}$ &  Temperature \\
	  \hline
	XXL-100 & Bias: & 7\% &  10\%  \\
	(11 clusters) & $\sigma$: & 50\% & 18\%\\
	XCS-DR1 & Bias: & 2\% &  5\%  \\
	(64 clusters)  & $\sigma$: & 55\% & 46\%\\
	 \hline
	 \label{tab:fit_xray}
    \end{tabular}
    \end{center}
\end{table}

\subsection{Nearby groups}
\label{sub:groups}
For the cosmological analysis for which this sample was constructed, the placing of on upper limit on the count-rate in the [0.5-2] keV band of 0.5 cts/s removed the majority of clusters below a redshift of $0.1$. The remaining clusters that have an assigned redshift of $z<0.1$ will most likely not be used in the cosmological analysis. The calculation of their X-ray properties highlighted some issues which seem to justify this decision. The cut in count-rate ensures that only very small groups are accepted into the original sample and as such they are extremely compact. This makes it difficult to disentangle any other possible sources of X-rays from either faint AGN, and/or occasionally the BCG of the cluster itself. These contribute to the 9\% of sources for which the X-ray property computations did not converge and these are marked with `**' in Table \ref{tab:catalogue}. In order to accurately measure the X-ray properties of these objects, one would need either deep XMM data to allow for spectral fitting or high resolution Chandra imaging to help with the removal of the contaminating point source or BCG.

\subsection{Distant clusters} 
\label{sub:distant}
As mentioned in Section \ref{sub:redshift_distribution}, we have a number of clusters for which we are unable to determine the redshift due to insufficient depth in the GROND data. Since the C1 selection of clusters is very pure, with only a minimal number of false detections, where we are unable to find a significant red sequence we assume that the cluster is distant. This assumption is supported by a number of observations of clusters already with either spectroscopic or photometric redshifts in the redshift range $0.9\la z\la 1.4$. Obtaining cluster photometric redshifts in this range has been shown to be feasible by \citet{Pierini2012}, where they studied the galaxy population of a single X-ray selected cluster at $z=1.1$ with data obtained from GROND. A separate program to obtain GROND photometric redshifts for some of these new detections lacking redshift information is currently underway with deeper observations and will form a useful sample for the study of high redshift clusters and their scaling relations in the future. 

\subsection{X-ray luminous clusters }
From Figure \ref{fig:z_Lx}, we are able to identify a subset of bright galaxy clusters with $L^{bol}_{500} > 5 \times 10^{44}\ \mathrm{erg\ s^{-1}}$ at redshifts $z>0.6$. The majority of these are already known and have been well studied and we find one new and potentially very interesting cluster. X-CLASS 2305, has no known counterpart in other cluster samples, including the Planck SZ cluster sample, despite having a luminosity $L^{bol}_{500}=1.2 \times 10^{45}\ \mathrm{erg\ s^{-1}}$. This cluster is subject of further study with Chandra and the Wide Field Imager (WFI), also on the MPG/ESO 2.2m telescope at La Silla (Clerc et al 2016, in prep). The already known clusters are: 
\begin{enumerate}
\item X-CLASS 228: This cluster is a part of XDCP with the alternate name XDCP J0954.2+1738 \citep{Nastasi2014}, where the bolometric luminosity is determined to be $L^{bol}_{500}=6.70\pm0.75 \times 10^{44}\ \mathrm{erg\ s^{-1}}$ in reasonable agreement with our value of $L^{bol}_{500}=5.68\times 10^{44}\ \mathrm{erg\ s^{-1}}$. Our measurement is probably affected by the presence of X-CLASS 229 which is located 2 \arcmin away. 
\item X-CLASS 439/440: This is a very well studied cluster with alternate names XMMXCS J015242.2-135746.8 and WARP J0152.7-1357 and it has been found in the ROSAT PSPC database by three independent groups \citep{Rosati1997rosat,ebeling2000warps,romer2000bright}. This is a difficult system to measure as it consists of two major components at $z=0.83$ and de-blending the emission from each of these components is difficult given that the separation of the two components are close together relative to the point-spread-function of XMM. 
\item X-CLASS 505: Another well studied cluster at $z=0.79$ also known as LCDCS 0504 \citep{Nelson2001, Johnson2006} and was the focus of a weak gravitational lensing analysis by \citet{guennou2014mass}.
\end{enumerate}

\section{Conclusions}
\label{sec:conclusions}
In this paper we present the first systematic followup of X-ray selected galaxy clusters with GROND along with a new method of determining photometric redshifts based on both optical and X-ray data simultaneously. We are able to confirm and provide redshifts for 236 out of 266 cluster candidates. Of these, 88 clusters were already spectroscopically confirmed and these provided a valuable set of targets on which the redshift algorithm could be tested and calibrated. Of the remaining clusters, 66 already had a photometric redshift available in the literature and we find that the accuracy of our measurement supersedes that of many of the previously published catalogues. The remainder of the clusters were previously unconfirmed cluster candidates and we report the first known redshifts for these objects. We find a median redshift of $z=0.39$ for this sample and report of photometric redshift accuracy of $\Delta z = 0.02 (1+z)$. 
We also present X-ray luminosities and temperatures and find a median bolometric luminosity of $4.6\times10^{44} \mathrm{erg\ s^{-1}}$ and a median temperature 2.6 keV. This sample of clusters will be used in a cosmological analysis following the z-CR-HR method of in a companion paper (Ridl et al., in prep). 
This survey can potentially carry on as long as XMM continues performing at its current levels and we expect and additional $\sim150$ clusters per year, $\sim50$ of which pass the cosmological selection criteria. Already, a second iteration of the X-ray detection pipeline on archival data up to January 2012 has produced 184 new cluster candidates. The methods presented here will also be useful for future studies with eROSITA, particularly in fields not falling into the footprints of existing wide-field optical surveys such as DES where pointed observations similar to these will be necessary to confirm cluster candidates and to obtain photometric redshifts. 
The catalogue is available at \url{http://xmm-lss.in2p3.fr:8080/l4sdb/}. 

\section*{Acknowledgements}
The authors thank F. Hofmann, M. Bernhardt,  G. Vasilopoulos and T. Schweyer and the support astronomers, A. Hempel, M. Rabus, R. Lachaume and I. Lacerna for additional help with the GROND observations. We also thank M. Pierre for the provision of valuable resources necessary for this project. JR thanks G. Erfanianfar, M. Mirkazemi, M. Klein, M. Jauzac and D. Pierini for useful discussions and J. Sanner for help with the visual inspection of photometric redshift results. This research made use of APLpy, an open-source plotting package for Python hosted at \url{http://aplpy.github.com} and Astropy, a community-developed core Python package for Astronomy \citep{astropy2013}. This research has made use of the NASA/IPAC Extragalactic Database (NED) which is operated by the Jet Propulsion Laboratory, California Institute of Technology, under contract with the National Aeronautics and Space Administration. Part of the funding for GROND (both hardware as well as personnel) was generously granted from the Leibniz-Prize to Prof. G. Hasinger (DFG grant HA 1850/28-1). TK and PW acknowledge support through the Sofja Kovalevskaja Award to P. Schady from the Alexander von Humboldt Foundation of Germany.




\bibliographystyle{mnras}
\bibliography{library2} 




\appendix
\section{The XCLASS/GROND observing program}
\label{app:observing}
This appendix provides an overview of the observing runs and the program of observations leading to the sample presented in this paper.

The observing campaigns were distributed over 6 semesters (P91 through P96). The program was designed to image X-CLASS galaxy cluster candidates without and with known redshift (calibration sample) and starting P93 it was extended to include targets that are outside the scope of this paper\footnote{X-CLASS sources detected on XMM archival pointings past April 2010 (Faccioli et al, in prep.)}.
Table~\ref{table:observ_runs} provides a summary of the observing runs, grouped by blocks of contiguous nights. In this table, observing nights of various quality and outcome are listed, regardless of the weather or technical conditions on site.

The GROND observation proposals were designed in order to achieve complete follow-up of the selected samples, taking into account weather and technical time losses inherited from previous runs. Most of the observing runs were allocated during dark time (critical for ensuring deep $g$ and $r$ band images). Time requests were calculated by considering that without interruption of the observing sequences, up to 20 X-CLASS fields and a few standard stars fields can be imaged during a 10-hour night. Compensation time was granted to account for interruptions due to ToO (target of opportunity) or instrument shutdown, resulting in a number of observed nights typically greater than the number of allocated nights in a given period.

Over the six observing semesters, the most significant changes impacting the observing schedule were: (i)~a failure in one of the two CCDs for each of the $i-$ and $z-$bands channels during P91; (ii)~a strong El~Ni{\~n}o event in 2015 affecting notably the P94, P95 and P96 semester observations resulting in an increased number of time losses due to bad weather conditions (wind, humidity and clouds)~; (iii)~recoating of the primary mirror (M1) in P95, resulting in a net improvement of the sensitivity of the telescope. 

In order to reach the depths and image quality required by the science objectives of the program, several targets were observed more than once and up to 8 times across the whole observing program. As described in Section \ref{sub:data_quality}, only the ``best" calibrated observing sequence was kept for the photometric redshift analysis of this paper.

The target lists for each observing run were established on the basis of visual inspection of the 3-colour and single-filter images acquired during previous runs. Whenever a dataset did not comply to the quality standards of the project, we added the corresponding target to the pool of objects requiring observations. These were then assigned priorities using a combination of empirical grades based on the image quality, observing night quality, seeing and limiting magnitude (for those fields that could be photometrically calibrated).

Observers were provided with prioritized target lists, finding charts and observation blocks (OBs), those accessible from the observation management tool P2PP. Observers were encouraged to select targets at high elevation\footnote{Usually taking advantage of the JSkyCalc software {\tt http://www.dartmouth.edu/~physics/labs/skycalc/flyer.html} to follow in real-time the availability of targets during an observing night.}, still accommodating for the on-site real-time observing conditions (e.g.~wind direction, atmospheric conditions, gamma-ray burst follow-up observations, etc.). At the end of each observing night a standardized log file was written, containing an entry for each OB that had been launched (time of observation, general conditions, comments). Selected entries in these observation logs can be made available upon request to the authors.

A typical X-CLASS/GROND observing night consists of: (i)~afternoon instrument calibration and preparation of the telescope~; (ii)~evening calibration (twilight flat fields) and standard fields acquisition~; (iii)~series of science OB and standard fields acquisition~and (iv)~morning calibration (twilight flat fields, biases, darks, etc.). Target of opportunity observations occurring during (iii) have a different ESO run identifier to those listed in Table~\ref{table:observ_runs}.

Finally, a typical X-CLASS/GROND science OB acquisition consists in: (i)~slewing the telescope to the target position~; (ii)~selecting a guide star on the guiding camera~; (iii)~launching the automated sequence of CCD/detectors integrations and readouts until completion of the observing block. Step (ii) has been the cause for repeated observations, due to the reduced availability of bright guide stars in the neighborhood of extragalactic science targets.

\begin{table*}
	\centering
\caption{\label{table:observ_runs}Table summary of the GROND observing campaign at the ESO/MPG-2.2m telescope relevant to the sample presented in this paper. The first column lists the standard run identifiers as referenced in the ESO archive database. The number of allocated nights takes into account target of opportunity (ToO) and technical overheads. These nights were also shared with separate programs to followup distant clusters as well as clusters from the updated X-ray processing, which are not included in this paper. The number of targets indicates the successful observations of XCLASS sources acquired during this period. The attachments between sources and observing runs is available through the L4SDataBase ({\tt http://xmm-lss.in2p3.fr:8080/l4sdb/}).}
		\begin{tabular}{@{}ccclllcl@{}}
\hline
ESO Run ID			&	Alloc.		&\multicolumn{3}{c}{Observation period (UT date at night start)} & N targets & Observers		\\
\hline
\hline
091.A-9017(A)		&	8	&	2013	&	Apr	&	7, 8				 	& 14	&	N. Clerc	\\
	"			&		&	2013	&	Aug	&	23, 24, 25, 26, 29		& 4	&	M.-L. Menzel	\\
092.A-9023(A)		&	12	&	2013	&	Oct	&	1, 2, 3, 4				& 29	&	N. Clerc\\
	"			&		&	2014	&	Jan	&	(fillers)				& 3	&	M. Salvato, F. Hofmann\\
	"			&		&	2014	&	Feb	&	26, 27, 28				& 21	&	J. Ridl, H. Steinle\\
	"			&		&	2014	&	Mar	&	1, 2, 3, 4, 5, 6, 7, 13		& 19	&	"			\\
093.A-9018(A)		&	16	&	2014	&	Apr	&	28, 29				& 2	&	J. Ridl	\\
	"			&		&	2014	&	May	&	1, 2, 3, 4, 5			& 9	&	"	\\
	"			&		&	2014	&	June	&	1, 5					& 2	&	Remote observing \\
	"			&		&	2014	&	Aug	&	24, 25, 26, 30, 31		& 9	&	M. Bernhardt, N. Clerc \\
	"			&		&	2014	&	Sep	&	2, 3					& 7	&	"\\
094.A-9018(A)		&	12	&	2014	&	Oct	&	16, 17, 18, 19, 20, 21	& 44	&	H. Steinle, G. Vasilopoulos	\\
	"			&		&	2014	&	Nov	&	12, 13, 14				& 17	&	"	\\	
	"			&		&	2015	&	Mar	&	11, 12, 13, 14, 15, 16, 17	& 27	&	H. Steinle, M. Salvato	\\
095.A-9008(A)		&	14	&	2015	&	Apr	&	16, 17, 18, 19			& 10	&	J. Ridl	\\
	"			&		&	2015	&	Sep	&	9, 10, 11, 12, 13, 16, 17	& 14	&	N. Clerc \\
096.A-9011(A)		&	14	&	2015	&	Nov	&	14, 15, 17, 28, 29, 30	& 2	&	J. Ridl, T. Schweyer	\\
	"			&		&	2015	&	Dec	&	13, 14, 15, 16			& 1	&	P. Wiseman	\\
	"			&		&	2016	&	Feb	&	5, 6, 7, 8, 9, 10			& 16	&	T. Kr{\"u}hler	\\
\hline
		\end{tabular}
\end{table*}

\section{The X-CLASS/GROND cluster catalogue}
\begin{table*}
\caption{The X-CLASS/GROND cluster catalogue.  }
 \label{tab:catalogue}
 \begin{tabular}{cccccccccc}
  \hline
ID & RA J2000& DEC J2000 & $z$  & $z_{lit}$ & $z$-Type & Count rate  & $r_{500}$ & $ L_{500}^{[0.5-2] keV}$ & $T_{L-T} $ \\ 
X-CLASS & (degrees) & (degrees) & GROND & Literature & Literature & (cts/s) & (Mpc) & ($10^{43}\mathrm{erg\ s^{-1}}$) & (keV)\\
(1) & (2) & (3) & (4) & (5) & (6) & (7) & (8) & (9) & (10) \\ \hline
20 & 193.438 & 10.195 & 0.63 & \  & \  & 0.034 & 0.7$\pm$0.2 & 5.6$\pm$0.8 & 3.5$\pm$0.8\\ 
35 & 196.274 & -10.279 & 0.34 & 0.330 & phot & 0.031 & 0.7$\pm$0.9 & 1.2$\pm$1.7 & 2.1$\pm$3.0\\ 
39 & 36.499 & -2.828 & 0.27 & 0.280 & conf & 0.014 & 0.5$\pm$0.1 & 0.3$\pm$0.1 & 1.4$\pm$0.3\\ 
40 & 35.189 & -3.434 & 0.32 & 0.330 & conf & 0.037 & 0.7$\pm$0.2 & 1.4$\pm$0.2 & 2.3$\pm$0.6\\ 
44 & 202.449 & 11.685 & 0.22 & \  & \  & 0.088 & 0.7$\pm$0.2 & 1.4$\pm$0.3 & 2.4$\pm$0.6\\ 
50 & 172.813 & -19.934 & 0.46 & \  & \  & 0.014 & 0.6$\pm$0.2 & 1.6$\pm$0.6 & 2.3$\pm$0.7\\ 
51 & 177.616 & 1.758 & F & \  & \  & 0.032 & * & * & *\\ 
54 & 145.938 & 16.738 & 0.16 & 0.180 & conf & 0.101 & 0.7$\pm$0.2 & 1.2$\pm$0.2 & 2.3$\pm$0.7\\ 
56 & 145.886 & 16.667 & 0.25 & 0.250 & conf & 0.187 & 0.9$\pm$0.2 & 3.7$\pm$0.3 & 3.4$\pm$0.9\\ 
57 & 145.995 & 16.688 & 0.32 & 0.250 & conf & 0.028 & 0.6$\pm$0.2 & 0.7$\pm$0.1 & 1.9$\pm$0.5\\ 
59 & 31.958 & 2.157 & D & \  & \  & 0.117 & * & * & *\\ 
65 & 339.252 & -15.273 & 0.31 & 0.300 & phot & 0.262 & 1.0$\pm$0.3 & 8.0$\pm$1.6 & 4.5$\pm$1.2\\ 
82 & 39.493 & -52.394 & 0.13 & 0.135 & conf & 0.215 & 0.8$\pm$0.2 & 1.2$\pm$0.1 & 2.3$\pm$0.6\\ 
86 & 348.766 & -58.935 & 0.44 & \  & \  & 0.020 & 0.6$\pm$0.9 & 1.4$\pm$2.0 & 2.2$\pm$3.1\\ 
87 & 349.095 & -59.076 & 0.62 & \  & \  & 0.048 & 0.8$\pm$0.2 & 7.5$\pm$1.4 & 4.0$\pm$1.1\\ 
88 & 183.395 & 2.896 & 0.36 & 0.410 & conf & 0.160 & 1.0$\pm$0.3 & 8.9$\pm$0.7 & 4.5$\pm$1.1\\ 
102 & 28.314 & 1.038 & 0.05 & 0.059 & conf & 0.354 & 0.6$\pm$0.1 & 0.3$\pm$0.1 & 1.4$\pm$0.4\\ 
135 & 300.803 & -32.798 & 0.28 & 0.260 & phot & 0.123 & 0.9$\pm$0.2 & 4.2$\pm$0.9 & 3.6$\pm$0.9\\ 
180 & 359.069 & -34.695 & D & \  & \  & 0.056 & * & * & *\\ 
205 & 314.089 & -4.630 & 0.54 & 0.583 & conf & 0.111 & 0.9$\pm$0.2 & 15.4$\pm$1.3 & 5.3$\pm$1.3\\ 
208 & 243.512 & -6.276 & 0.49 & \  & \  & 0.026 & 0.8$\pm$1.1 & 4.5$\pm$6.5 & 3.4$\pm$4.8\\ 
219 & 190.801 & 13.220 & 0.80 & 0.791 & phot & 0.046 & 0.7$\pm$0.2 & 11.6$\pm$2.0 & 4.4$\pm$1.1\\ 
224 & 36.377 & -4.240 & 0.13 & 0.140 & conf & 0.243 & 0.8$\pm$0.2 & 1.4$\pm$0.2 & 2.5$\pm$0.7\\ 
228 & 148.572 & 17.634 & 0.83 & 0.828 & phot & 0.084 & 0.8$\pm$0.2 & 22.3$\pm$3.0 & 5.5$\pm$1.4\\ 
229 & 148.582 & 17.597 & 0.40 & 0.380 & conf & 0.127 & 0.9$\pm$0.3 & 6.0$\pm$1.6 & 4.0$\pm$1.2\\ 
233 & 10.729 & -18.011 & 0.24 & \  & \  & 0.015 & 0.5$\pm$0.7 & 0.2$\pm$0.2 & 1.1$\pm$1.5\\ 
237 & 0.270 & -25.066 & D & 0.910 & phot & 0.021 & 0.6$\pm$0.2 & 8.0$\pm$1.4 & 3.6$\pm$0.8\\ 
238 & 0.125 & -25.203 & 0.13 & 0.150 & conf & 0.107 & 0.7$\pm$0.2 & 0.7$\pm$0.1 & 1.9$\pm$0.5\\ 
244 & 21.394 & -1.279 & 0.59 & 0.490 & phot & 0.030 & 0.7$\pm$0.2 & 4.7$\pm$0.7 & 3.4$\pm$0.8\\ 
245 & 21.402 & -1.431 & 0.14 & 0.019 & conf & 0.151 & 1.7$\pm$0.5 & 42.2$\pm$19.0 & 9.3$\pm$2.9\\ 
263 & 213.741 & -0.349 & 0.12 & 0.140 & conf & 0.149 & 0.7$\pm$0.2 & 0.7$\pm$0.1 & 1.9$\pm$0.5\\ 
270 & 353.083 & 19.917 & 0.26 & \  & \  & 0.033 & 0.7$\pm$0.2 & 1.2$\pm$0.3 & 2.2$\pm$0.6\\ 
287 & 358.069 & -26.093 & 0.25 & 0.275 & tent & 0.044 & 0.6$\pm$0.6 & 0.7$\pm$0.7 & 1.8$\pm$1.8\\ 
300 & 53.620 & -36.238 & 0.33 & \  & \  & 0.034 & 0.7$\pm$0.2 & 1.1$\pm$0.2 & 2.1$\pm$0.5\\ 
314 & 56.257 & -41.213 & 0.44 & \  & \  & 0.144 & 1.0$\pm$0.2 & 9.5$\pm$1.6 & 4.6$\pm$1.1\\ 
335 & 35.287 & 19.968 & 0.44 & 0.450 & phot & 0.223 & 1.1$\pm$0.3 & 21.6$\pm$3.1 & 6.3$\pm$1.6\\ 
372 & 45.526 & -0.001 & 0.68 & 0.340 & tent & 0.030 & 0.7$\pm$0.2 & 6.6$\pm$1.2 & 3.7$\pm$1.0\\ 
374 & 177.549 & 1.646 & 0.37 & 0.450 & phot & 0.049 & 0.7$\pm$0.2 & 1.9$\pm$0.5 & 2.6$\pm$0.7\\ 
377 & 6.648 & 17.159 & 0.34 & 0.390 & conf & 0.289 & 1.1$\pm$0.3 & 15.1$\pm$0.8 & 5.6$\pm$1.4\\ 
378 & 6.708 & 17.325 & 0.47 & 0.491 & conf & 0.025 & 0.7$\pm$0.2 & 2.6$\pm$0.4 & 2.8$\pm$0.7\\ 
382 & 180.204 & -3.458 & 0.39 & 0.396 & phot & 0.179 & 1.0$\pm$0.3 & 9.0$\pm$1.4 & 4.6$\pm$1.2\\ 
386 & 193.143 & -29.417 & 0.25 & \  & \  & 0.018 & 0.5$\pm$0.1 & 0.4$\pm$0.1 & 1.4$\pm$0.3\\ 
387 & 193.227 & -29.456 & D & 1.240 & conf & 0.030 & ** & ** & **\\ 
399 & 170.958 & 5.496 & 0.62 & 0.650 & conf & 0.045 & 0.8$\pm$0.2 & 7.8$\pm$0.8 & 4.0$\pm$1.0\\ 
407 & 208.943 & 18.382 & 0.36 & 0.290 & phot & 0.026 & 0.7$\pm$0.2 & 1.2$\pm$0.2 & 2.2$\pm$0.6\\ 
408 & 59.354 & 1.300 & 0.23 & 0.130 & phot & 0.096 & ** & ** & **\\ 
412 & 164.104 & -3.589 & 0.66 & 0.630 & phot & 0.086 & 0.9$\pm$0.2 & 15.2$\pm$1.8 & 5.1$\pm$1.2\\ 
414 & 210.317 & 2.752 & 0.24 & 0.238 & phot & 0.044 & 0.7$\pm$0.2 & 0.8$\pm$0.2 & 1.9$\pm$0.5\\ 
417 & 39.136 & -52.392 & 0.60 & \  & \  & 0.018 & 0.6$\pm$0.2 & 2.8$\pm$0.5 & 2.7$\pm$0.7\\ 
418 & 39.022 & -52.421 & 0.59 & \  & \  & 0.045 & 0.8$\pm$0.2 & 6.2$\pm$1.0 & 3.7$\pm$0.9\\ 
419 & 337.096 & -5.342 & 0.39 & 0.350 & phot & 0.046 & 0.8$\pm$0.2 & 2.8$\pm$0.5 & 3.0$\pm$0.8\\ 
420 & 155.739 & 19.886 & 0.81 & \  & \  & 0.011 & 0.6$\pm$0.1 & 3.4$\pm$0.7 & 2.7$\pm$0.6\\ 
424 & 333.903 & -17.760 & 0.41 & 0.400 & phot & 0.049 & 0.8$\pm$0.2 & 2.8$\pm$0.6 & 2.9$\pm$0.8\\ 
430 & 54.438 & -25.378 & 0.53 & 0.585 & conf & 0.040 & 0.8$\pm$0.2 & 5.5$\pm$0.6 & 3.5$\pm$0.9\\ 
435 & 156.003 & 4.038 & 0.47 & 0.480 & phot & 0.012 & 0.6$\pm$0.2 & 1.0$\pm$0.2 & 2.0$\pm$0.5\\ 
439 & 28.187 & -13.953 & 0.84 & 0.831 & phot & 0.182 & 0.9$\pm$0.3 & 32.3$\pm$5.4 & 6.4$\pm$1.7\\ 
440 & 28.166 & -13.975 & 0.84 & 0.831 & phot & 0.085 & 0.8$\pm$0.2 & 24.2$\pm$2.4 & 5.7$\pm$1.4\\ 
\hline
 \end{tabular}
 \raggedright{\ \ \ \ \ \ In column 4: the flag `F' indicates that we were unable to obtain a secure redshift from the GROND observations as discussed in Section \ref{sub:unconfirmed} and `D' that the cluster has be classified as distant.\\
 In columns 8-10: * indicates that we were unable to compute X-ray properties due to the lack of a secure redshift and ** that the X-ray processing pipeline failed to converge on a reasonable value.}

\end{table*}
\begin{table*}
 \contcaption{The X-CLASS/GROND catalogue}
 \label{tab:continued}
 \begin{tabular}{cccccccccc}
  \hline
  ID &RA J2000& DEC J2000 & $z$  & $z_{lit}$ & $z$-Type & Count rate  & $r_{500}$ & $ L_{500}^{[0.5-2] keV}$ & $T_{L-T} $ \\ 
X-CLASS & (degrees) & (degrees) & GROND & Literature & Literature & (cts/s) & (Mpc) & ($10^{43}\mathrm{erg\ s^{-1}}$) & (keV)\\
(1) & (2) & (3) & (4) & (5) & (6) & (7) & (8) & (9) & (10) \\ \hline
441 & 28.090 & -14.087 & 0.32 & \  & \  & 0.052 & 0.7$\pm$0.2 & 1.7$\pm$0.4 & 2.5$\pm$0.6\\ 
442 & 28.241 & -14.114 & 0.67 & 0.745 & conf & 0.048 & 0.8$\pm$0.2 & 10.2$\pm$1.6 & 4.2$\pm$1.1\\ 
453 & 191.230 & -0.445 & 0.23 & 0.220 & tent & 0.031 & 0.6$\pm$0.2 & 0.5$\pm$0.2 & 1.6$\pm$0.5\\ 
454 & 191.225 & -0.559 & 0.22 & 0.230 & conf & 0.045 & 0.6$\pm$0.2 & 0.7$\pm$0.1 & 1.9$\pm$0.5\\ 
459 & 4.572 & 16.294 & 0.55 & 0.550 & conf & 0.098 & 0.9$\pm$0.2 & 11.5$\pm$0.7 & 4.8$\pm$1.2\\ 
462 & 76.332 & -28.815 & 0.46 & 0.509 & conf & 0.050 & 0.8$\pm$0.2 & 4.5$\pm$0.5 & 3.4$\pm$0.9\\ 
469 & 202.662 & -1.643 & D & 0.660 & tent & 0.013 & ** & ** & **\\ 
470 & 208.572 & -2.366 & 0.53 & 0.546 & conf & 0.100 & 0.9$\pm$0.2 & 11.8$\pm$1.0 & 4.8$\pm$1.2\\ 
476 & 36.859 & -4.538 & 0.32 & 0.307 & conf & 0.021 & 0.6$\pm$0.2 & 0.8$\pm$0.2 & 1.9$\pm$0.5\\ 
477 & 36.353 & -4.680 & 0.29 & 0.266 & conf & 0.091 & 0.8$\pm$0.2 & 2.1$\pm$0.2 & 2.8$\pm$0.7\\ 
478 & 173.116 & -34.568 & 0.60 & \  & \  & 0.011 & 0.6$\pm$0.2 & 2.1$\pm$0.3 & 2.5$\pm$0.6\\ 
479 & 173.133 & -34.731 & 0.53 & \  & \  & 0.084 & 0.9$\pm$0.2 & 10.1$\pm$1.7 & 4.6$\pm$1.2\\ 
485 & 161.182 & -1.332 & D & 0.750 & tent & 0.016 & 0.6$\pm$0.1 & 6.6$\pm$2.0 & 3.4$\pm$0.8\\ 
495 & 151.960 & 12.972 & D & 1.082 & conf & 0.010 & 0.5$\pm$0.1 & 6.4$\pm$1.3 & 3.2$\pm$0.6\\ 
499 & 65.073 & -50.532 & 0.39 & \  & \  & 0.066 & 0.8$\pm$0.2 & 3.6$\pm$0.5 & 3.2$\pm$0.8\\ 
500 & 46.561 & -0.095 & 0.36 & 0.430 & conf & 0.261 & 0.9$\pm$0.3 & 8.3$\pm$2.9 & 4.4$\pm$1.4\\ 
501 & 46.573 & -0.141 & 0.12 & 0.109 & conf & 0.195 & 0.7$\pm$0.2 & 0.7$\pm$0.2 & 1.9$\pm$0.5\\ 
502 & 184.169 & -12.074 & 0.68 & 0.790 & tent & 0.085 & 0.8$\pm$0.2 & 10.1$\pm$2.4 & 4.3$\pm$1.2\\ 
503 & 184.109 & -11.962 & 0.60 & \  & \  & 0.016 & 0.7$\pm$0.2 & 2.9$\pm$0.6 & 2.8$\pm$0.7\\ 
505 & 184.190 & -12.022 & D & 0.794 & conf & 0.101 & 0.9$\pm$0.2 & 25.6$\pm$2.0 & 5.9$\pm$1.5\\ 
507 & 1.000 & -35.948 & 0.51 & \  & \  & 0.041 & 0.7$\pm$0.2 & 3.6$\pm$0.7 & 3.1$\pm$0.8\\ 
510 & 17.010 & -80.311 & 0.34 & \  & \  & 0.066 & 0.8$\pm$0.2 & 2.9$\pm$0.8 & 3.0$\pm$0.9\\ 
514 & 42.529 & -31.067 & D & 0.910 & conf & 0.047 & 0.7$\pm$0.2 & 17.2$\pm$1.8 & 4.9$\pm$1.2\\ 
517 & 351.397 & -11.994 & 0.40 & \  & \  & 0.019 & 0.6$\pm$0.1 & 1.0$\pm$0.2 & 2.0$\pm$0.5\\ 
527 & 222.539 & 9.075 & 0.58 & 0.640 & conf & 0.031 & 0.8$\pm$0.2 & 6.8$\pm$1.3 & 3.8$\pm$1.0\\ 
528 & 73.587 & -53.259 & 0.43 & \  & \  & 0.029 & ** & ** & **\\ 
530 & 73.779 & -53.399 & 0.41 & 0.410 & conf & 0.060 & ** & ** & **\\ 
531 & 8.949 & -43.379 & 0.62 & 0.630 & conf & 0.017 & 0.7$\pm$0.2 & 3.2$\pm$0.4 & 2.9$\pm$0.7\\ 
533 & 8.616 & -43.316 & 0.42 & 0.390 & conf & 0.196 & 1.0$\pm$0.3 & 9.7$\pm$0.5 & 4.7$\pm$1.2\\ 
534 & 8.443 & -43.292 & 0.22 & 0.220 & conf & 0.149 & 0.8$\pm$0.2 & 2.2$\pm$0.2 & 2.8$\pm$0.7\\ 
536 & 339.853 & -5.788 & 0.26 & 0.242 & phot & 0.317 & 1.0$\pm$0.3 & 6.6$\pm$1.5 & 4.3$\pm$1.2\\ 
538 & 339.892 & -6.006 & 0.10 & 0.173 & phot & 0.055 & 0.5$\pm$0.2 & 0.2$\pm$0.1 & 1.2$\pm$0.4\\ 
540 & 341.195 & -72.736 & 0.19 & \  & \  & 0.028 & 0.5$\pm$0.1 & 0.3$\pm$0.1 & 1.3$\pm$0.3\\ 
541 & 341.492 & -72.748 & 0.09 & \  & \  & 0.203 & 0.6$\pm$0.2 & 0.5$\pm$0.2 & 1.7$\pm$0.5\\ 
542 & 223.322 & 3.578 & 0.33 & 0.346 & phot & 0.056 & 0.8$\pm$0.2 & 2.5$\pm$0.4 & 2.9$\pm$0.8\\ 
551 & 5.619 & -48.726 & D & \  & \  & 0.023 & * & * & *\\ 
553 & 198.731 & -16.642 & 0.69 & 0.610 & phot & 0.034 & 0.7$\pm$0.2 & 7.1$\pm$1.6 & 3.8$\pm$1.1\\ 
560 & 195.647 & -2.309 & D & 0.620 & tent & 0.012 & 0.7$\pm$0.7 & 3.2$\pm$3.4 & 2.9$\pm$3.0\\ 
562 & 229.102 & -0.832 & 0.42 & 0.380 & tent & 0.103 & 0.9$\pm$0.2 & 7.4$\pm$1.2 & 4.2$\pm$1.1\\ 
567 & 229.243 & -1.111 & 0.12 & 0.117 & conf & 0.226 & 0.8$\pm$0.2 & 1.3$\pm$0.2 & 2.4$\pm$0.7\\ 
569 & 312.031 & -17.699 & 0.17 & \  & \  & 0.101 & 0.8$\pm$0.2 & 1.3$\pm$0.5 & 2.4$\pm$0.8\\ 
634 & 49.572 & -3.035 & 0.41 & 0.370 & phot & 0.107 & 0.9$\pm$0.2 & 7.0$\pm$1.2 & 4.1$\pm$1.0\\ 
872 & 156.213 & -18.563 & D & \  & \  & 0.061 & * & * & *\\ 
890 & 20.273 & 3.802 & 0.35 & 0.340 & phot & 0.113 & 0.9$\pm$0.2 & 5.0$\pm$1.0 & 3.7$\pm$1.0\\ 
911 & 78.082 & -32.747 & 0.61 & \  & \  & 0.039 & 0.7$\pm$0.2 & 5.7$\pm$1.2 & 3.6$\pm$1.0\\ 
924 & 45.813 & 16.438 & 0.04 & 0.032 & tent & 0.042 & ** & ** & **\\ 
927 & 12.418 & -29.588 & 0.35 & 0.108 & tent & 0.043 & 0.8$\pm$0.2 & 2.5$\pm$0.9 & 2.9$\pm$0.9\\ 
955 & 2.206 & -32.264 & 0.18 & 0.267 & tent & 0.029 & 0.6$\pm$0.2 & 0.3$\pm$0.1 & 1.4$\pm$0.4\\ 
964 & 234.184 & -14.173 & 0.40 & 0.400 & conf & 0.312 & 1.2$\pm$0.3 & 24.8$\pm$2.0 & 6.7$\pm$1.6\\ 
967 & 310.411 & -35.147 & 0.41 & 0.430 & conf & 0.125 & 0.9$\pm$0.2 & 8.9$\pm$1.0 & 4.5$\pm$1.1\\ 
996 & 195.731 & -15.677 & F & \  & \  & 0.020 & * & * & *\\ 
997 & 195.715 & -15.701 & F & \  & \  & 0.044 & * & * & *\\ 
998 & 195.582 & -15.718 & F & \  & \  & 0.014 & * & * & *\\ 
1014 & 30.240 & -9.354 & 0.31 & 0.338 & tent & 0.032 & 0.7$\pm$0.2 & 1.7$\pm$0.6 & 2.5$\pm$0.8\\ 
1030 & 3.368 & -27.379 & 0.40 & \  & \  & 0.157 & 1.0$\pm$0.2 & 8.4$\pm$1.4 & 4.5$\pm$1.1\\ 
1032 & 149.887 & 5.428 & 0.24 & 0.250 & phot & 0.011 & 0.5$\pm$0.1 & 0.3$\pm$0.1 & 1.3$\pm$0.3\\ 
1059 & 358.902 & 5.855 & 0.27 & 0.280 & phot & 0.050 & 0.7$\pm$0.2 & 1.6$\pm$0.4 & 2.5$\pm$0.7\\ 
1117 & 40.097 & -23.289 & D & \  & \  & 0.016 & * & * & *\\ 
1125 & 162.402 & -13.968 & 0.36 & \  & \  & 0.050 & 0.8$\pm$0.2 & 2.7$\pm$0.6 & 3.0$\pm$0.8\\ 
1126 & 162.698 & -14.172 & 0.53 & \  & \  & 0.087 & 0.9$\pm$0.2 & 9.1$\pm$1.0 & 4.4$\pm$1.1\\ 
1146 & 335.062 & -28.044 & 0.36 & 0.165 & tent & 0.029 & 0.7$\pm$0.2 & 1.2$\pm$0.3 & 2.2$\pm$0.6\\ 
1195 & 323.419 & -0.643 & 0.23 & 0.211 & tent & 0.172 & 0.9$\pm$0.2 & 3.2$\pm$0.8 & 3.3$\pm$0.9\\
\hline
 \end{tabular}
\end{table*}
\begin{table*}
 \contcaption{The X-CLASS/GROND catalogue}
 \label{tab:continued}
 \begin{tabular}{cccccccccc}
  \hline
  ID & RA J2000& DEC J2000 & $z$  & $z_{lit}$ & $z$-Type & Count rate  & $r_{500}$ & $ L_{500}^{[0.5-2] keV}$ & $T_{L-T} $ \\ 
X-CLASS & (degrees) & (degrees) & GROND & Literature & Literature & (cts/s) & (Mpc) & ($10^{43}\mathrm{erg\ s^{-1}}$) & (keV)\\
(1) & (2) & (3) & (4) & (5) & (6) & (7) & (8) & (9) & (10) \\ \hline
1218 & 37.440 & -29.631 & 0.06 & 0.061 & conf & 0.143 & 0.5$\pm$0.1 & 0.1$\pm$0.0 & 1.0$\pm$0.3\\ 
1219 & 174.013 & -3.497 & 0.80 & \  & \  & 0.054 & 0.8$\pm$0.2 & 14.3$\pm$2.5 & 4.7$\pm$1.2\\ 
1239 & 218.691 & -32.686 & 0.08 & 0.087 & tent & 0.166 & 0.6$\pm$0.2 & 0.4$\pm$0.2 & 1.6$\pm$0.6\\ 
1296 & 92.046 & -61.896 & 0.24 & \  & \  & 0.068 & 0.7$\pm$0.2 & 1.3$\pm$0.3 & 2.3$\pm$0.6\\ 
1297 & 91.901 & -61.928 & 0.33 & \  & \  & 0.242 & 1.0$\pm$0.3 & 9.3$\pm$1.5 & 4.8$\pm$1.2\\ 
1345 & 125.398 & 1.042 & 0.09 & 0.130 & phot & 0.353 & 0.8$\pm$0.2 & 1.3$\pm$0.6 & 2.4$\pm$0.8\\ 
1352 & 51.157 & -3.190 & 0.52 & \  & \  & 0.047 & 0.8$\pm$0.2 & 5.0$\pm$1.7 & 3.5$\pm$1.0\\ 
1386 & 17.576 & 19.638 & 0.32 & 0.317 & conf & 0.081 & 0.8$\pm$0.2 & 2.9$\pm$0.4 & 3.1$\pm$0.8\\ 
1400 & 63.674 & 14.447 & F & \  & \  & 0.047 & * & * & *\\ 
1424 & 215.314 & 3.130 & 0.19 & 0.310 & phot & 0.069 & 0.7$\pm$0.2 & 0.8$\pm$0.3 & 2.0$\pm$0.6\\ 
1425 & 322.662 & 4.919 & 0.61 & \  & \  & 0.051 & 0.8$\pm$0.2 & 8.3$\pm$1.3 & 4.1$\pm$1.0\\ 
1449 & 13.250 & -8.661 & 0.32 & 0.315 & conf & 0.073 & 0.8$\pm$0.2 & 3.5$\pm$0.5 & 3.3$\pm$0.8\\ 
1478 & 352.180 & -55.567 & 0.60 & 0.830 & phot & 0.085 & 0.9$\pm$0.2 & 12.9$\pm$2.3 & 4.9$\pm$1.3\\ 
1480 & 349.822 & -55.326 & 0.16 & 0.180 & phot & 0.103 & 0.7$\pm$0.2 & 0.8$\pm$0.2 & 1.9$\pm$0.5\\ 
1482 & 349.222 & -54.906 & 0.38 & 0.440 & phot & 0.255 & 1.0$\pm$0.3 & 12.2$\pm$2.2 & 5.2$\pm$1.4\\ 
1483 & 351.639 & -55.022 & 0.41 & 0.320 & phot & 0.064 & 0.8$\pm$0.2 & 3.8$\pm$0.6 & 3.3$\pm$0.8\\ 
1485 & 352.008 & -54.929 & D & 0.960 & phot & 0.025 & 0.6$\pm$0.2 & 10.4$\pm$2.3 & 4.0$\pm$1.0\\ 
1486 & 349.934 & -54.640 & 0.52 & 0.550 & phot & 0.019 & 0.7$\pm$0.2 & 2.5$\pm$0.4 & 2.7$\pm$0.7\\ 
1487 & 351.396 & -54.723 & 0.15 & 0.169 & phot & 0.064 & 0.6$\pm$0.2 & 0.5$\pm$0.2 & 1.6$\pm$0.5\\ 
1488 & 352.502 & -54.619 & 0.18 & 0.176 & conf & 0.229 & 0.8$\pm$0.2 & 2.1$\pm$0.2 & 2.8$\pm$0.7\\ 
1489 & 352.418 & -54.790 & 0.15 & 0.139 & phot & 0.021 & 0.0$\pm$0.4 & 0.0$\pm$0.1 & 0.0$\pm$0.8\\ 
1490 & 353.884 & -54.588 & D & 0.670 & phot & 0.030 & 0.8$\pm$0.2 & 7.4$\pm$1.3 & 3.9$\pm$1.0\\ 
1581 & 148.809 & 18.208 & 0.42 & 0.416 & conf & 0.018 & 0.7$\pm$0.2 & 1.9$\pm$0.5 & 2.5$\pm$0.7\\ 
1582 & 148.814 & 18.062 & 0.65 & \  & \  & 0.024 & 0.7$\pm$0.2 & 5.3$\pm$1.2 & 3.4$\pm$0.9\\ 
1620 & 86.796 & -51.202 & 0.26 & \  & \  & 0.072 & 0.7$\pm$0.2 & 1.7$\pm$0.3 & 2.5$\pm$0.7\\ 
1688 & 26.205 & -4.550 & 0.14 & 0.170 & phot & 0.037 & 0.5$\pm$0.2 & 0.2$\pm$0.1 & 1.3$\pm$0.4\\ 
1691 & 60.056 & -67.599 & 0.52 & \  & \  & 0.054 & 0.8$\pm$0.2 & 6.1$\pm$0.8 & 3.8$\pm$0.9\\ 
1693 & 59.765 & -67.727 & 0.05 & 0.070 & tent & 0.030 & 0.4$\pm$0.1 & 0.0$\pm$0.0 & 0.6$\pm$0.2\\ 
1705 & 34.636 & -5.016 & D & 0.880 & conf & 0.014 & 0.6$\pm$0.2 & 5.2$\pm$0.7 & 3.1$\pm$0.7\\ 
1706 & 34.938 & -4.891 & 0.35 & 0.330 & conf & 0.019 & 0.6$\pm$0.2 & 0.7$\pm$0.1 & 1.8$\pm$0.5\\ 
1773 & 341.460 & -52.912 & 0.45 & \  & \  & 0.091 & 0.9$\pm$0.2 & 5.9$\pm$0.9 & 3.8$\pm$0.9\\ 
1801 & 332.777 & -16.950 & 0.31 & \  & \  & 0.045 & 0.7$\pm$0.2 & 1.7$\pm$0.5 & 2.5$\pm$0.7\\ 
1809 & 302.081 & -44.595 & 0.52 & \  & \  & 0.023 & 0.7$\pm$0.2 & 2.8$\pm$0.5 & 2.8$\pm$0.7\\ 
1811 & 36.870 & -40.852 & 0.42 & 0.400 & tent & 0.136 & 0.9$\pm$0.2 & 8.0$\pm$1.4 & 4.4$\pm$1.1\\ 
1814 & 5.404 & -8.604 & 0.36 & \  & \  & 0.022 & 0.6$\pm$0.2 & 1.0$\pm$0.2 & 2.0$\pm$0.5\\ 
1818 & 37.959 & -7.477 & 0.11 & 0.179 & phot & 0.029 & 0.5$\pm$0.2 & 0.1$\pm$0.1 & 1.0$\pm$0.4\\ 
1819 & 32.553 & -0.247 & 0.30 & 0.280 & phot & 0.020 & 0.6$\pm$0.2 & 0.7$\pm$0.2 & 1.8$\pm$0.5\\ 
1821 & 52.263 & 2.940 & 0.35 & 0.410 & conf & 0.040 & 0.7$\pm$0.2 & 2.8$\pm$0.3 & 2.9$\pm$0.7\\ 
1827 & 9.368 & -33.890 & 0.36 & 0.348 & tent & 0.072 & 0.8$\pm$0.2 & 3.2$\pm$0.5 & 3.1$\pm$0.8\\ 
1837 & 163.600 & -11.774 & 0.55 & 0.700 & conf & 0.018 & 0.6$\pm$0.2 & 3.9$\pm$0.6 & 3.0$\pm$0.8\\ 
1838 & 163.488 & -11.816 & 0.68 & \  & \  & 0.014 & 0.6$\pm$0.2 & 3.0$\pm$0.8 & 2.7$\pm$0.8\\ 
1845 & 334.410 & -35.867 & 0.85 & \  & \  & 0.026 & 0.7$\pm$0.7 & 7.2$\pm$7.3 & 3.6$\pm$3.6\\ 
1851 & 33.473 & -73.921 & 0.43 & \  & \  & 0.015 & 0.6$\pm$0.2 & 1.5$\pm$0.3 & 2.3$\pm$0.6\\ 
1853 & 350.358 & 19.753 & 0.30 & 0.400 & phot & 0.239 & 1.0$\pm$0.2 & 7.1$\pm$1.3 & 4.3$\pm$1.1\\ 
1854 & 350.535 & 19.730 & 0.53 & 0.500 & phot & 0.065 & 0.8$\pm$0.3 & 6.2$\pm$2.3 & 3.8$\pm$1.3\\ 
1855 & 350.588 & 19.647 & 0.21 & 0.230 & phot & 0.038 & 0.6$\pm$0.2 & 0.5$\pm$0.1 & 1.6$\pm$0.4\\ 
1856 & 3.862 & 17.290 & 0.47 & \  & \  & 0.030 & 0.7$\pm$0.2 & 2.6$\pm$0.5 & 2.8$\pm$0.7\\ 
1858 & 205.771 & -0.015 & 0.70 & 0.600 & phot & 0.124 & 0.9$\pm$0.2 & 22.6$\pm$3.0 & 5.8$\pm$1.5\\ 
1862 & 190.793 & 14.340 & 0.37 & 0.340 & conf & 0.037 & 0.7$\pm$0.2 & 1.8$\pm$0.3 & 2.6$\pm$0.7\\ 
1864 & 130.351 & 0.774 & 0.41 & 0.410 & conf & 0.043 & 0.7$\pm$0.2 & 2.7$\pm$0.4 & 2.9$\pm$0.8\\ 
1868 & 358.469 & -15.217 & 0.52 & \  & \  & 0.025 & 0.7$\pm$0.2 & 2.5$\pm$0.4 & 2.7$\pm$0.6\\ 
1874 & 150.423 & 2.425 & 0.13 & 0.120 & conf & 0.115 & 0.6$\pm$0.2 & 0.5$\pm$0.1 & 1.6$\pm$0.4\\ 
1876 & 150.507 & 2.226 & 0.84 & 0.830 & conf & 0.044 & 0.7$\pm$0.2 & 12.7$\pm$1.3 & 4.5$\pm$1.1\\ 
1877 & 150.125 & 2.696 & 0.35 & 0.350 & phot & 0.039 & 0.7$\pm$0.2 & 1.8$\pm$0.3 & 2.5$\pm$0.7\\ 
1879 & 150.058 & 2.379 & 0.32 & 0.350 & conf & 0.027 & 0.6$\pm$0.2 & 1.0$\pm$0.3 & 2.0$\pm$0.7\\ 
1880 & 150.093 & 2.391 & 0.23 & 0.220 & conf & 0.013 & 0.6$\pm$0.2 & 0.3$\pm$0.1 & 1.4$\pm$0.4\\ 
1882 & 150.196 & 1.658 & 0.22 & 0.220 & conf & 0.189 & 0.9$\pm$0.2 & 2.8$\pm$0.2 & 3.1$\pm$0.8\\ 
1883 & 150.182 & 1.768 & 0.34 & 0.350 & conf & 0.021 & 0.6$\pm$0.2 & 1.0$\pm$0.2 & 2.0$\pm$0.6\\ 
1886 & 150.030 & 2.209 & D & 0.930 & conf & 0.010 & 0.6$\pm$0.2 & 5.0$\pm$1.1 & 3.0$\pm$0.8\\ 
1888 & 149.600 & 2.820 & 0.35 & 0.340 & conf & 0.036 & 0.7$\pm$0.2 & 1.6$\pm$0.3 & 2.5$\pm$0.7\\ 
1889 & 134.606 & 13.958 & 0.49 & 0.488 & phot & 0.057 & 0.8$\pm$0.2 & 5.3$\pm$0.8 & 3.6$\pm$0.9\\ 
1892 & 5.416 & -15.075 & 0.56 & \  & \  & 0.064 & 0.8$\pm$0.2 & 7.0$\pm$1.0 & 3.9$\pm$1.0\\ 
\hline
 \end{tabular}
\end{table*}
\begin{table*}
 \contcaption{The X-CLASS/GROND catalogue}
 \label{tab:continued}
 \begin{tabular}{cccccccccc}
  \hline 
  ID & RA J2000& DEC J2000 & $z$  & $z_{lit}$ & $z$-Type & Count rate  & $r_{500}$ & $ L_{500}^{[0.5-2] keV}$ & $T_{L-T} $ \\ 
X-CLASS & (degrees) & (degrees) & GROND & Literature & Literature & (cts/s) & (Mpc) & ($10^{43}\mathrm{erg\ s^{-1}}$) & (keV)\\
(1) & (2) & (3) & (4) & (5) & (6) & (7) & (8) & (9) & (10) \\ \hline
1893 & 5.559 & -15.098 & 0.53 & \  & \  & 0.028 & 0.7$\pm$0.2 & 3.0$\pm$0.4 & 2.9$\pm$0.7\\ 
1896 & 169.360 & 7.727 & 0.48 & 0.480 & conf & 0.086 & 0.9$\pm$0.3 & 7.1$\pm$2.0 & 4.1$\pm$1.3\\ 
1900 & 9.843 & 0.802 & 0.36 & 0.410 & conf & 0.041 & 0.7$\pm$0.2 & 2.7$\pm$0.4 & 2.9$\pm$0.8\\ 
1903 & 67.148 & -17.146 & 0.84 & \  & \  & 0.020 & 0.6$\pm$0.2 & 6.0$\pm$0.8 & 3.4$\pm$0.9\\ 
1906 & 328.656 & -9.261 & 0.08 & 0.078 & conf & 0.170 & 0.6$\pm$0.1 & 0.3$\pm$0.0 & 1.4$\pm$0.3\\ 
1908 & 37.778 & -54.064 & 0.56 & \  & \  & 0.154 & 1.0$\pm$0.3 & 16.4$\pm$2.1 & 5.4$\pm$1.4\\ 
1928 & 73.502 & -3.143 & 0.26 & 0.260 & tent & 0.081 & 0.8$\pm$0.2 & 2.0$\pm$0.5 & 2.7$\pm$0.7\\ 
1943 & 149.162 & -0.360 & 0.03 & 0.087 & conf & 0.348 & 0.7$\pm$0.2 & 0.7$\pm$0.1 & 2.0$\pm$0.5\\ 
1944 & 149.044 & -0.365 & 0.57 & 0.580 & phot & 0.039 & 0.8$\pm$0.2 & 5.3$\pm$1.0 & 3.5$\pm$0.9\\ 
1954 & 54.353 & -34.955 & D & 0.840 & conf & 0.021 & 0.6$\pm$0.2 & 6.3$\pm$0.4 & 3.4$\pm$0.8\\ 
1955 & 36.017 & -4.226 & 0.24 & 1.050 & conf & 0.039 & 0.7$\pm$0.2 & 19.1$\pm$1.6 & 4.8$\pm$1.2\\ 
1956 & 36.146 & -4.249 & 0.24 & 0.262 & conf & 0.038 & 0.6$\pm$0.2 & 0.8$\pm$0.1 & 1.9$\pm$0.5\\ 
1992 & 149.921 & 2.521 & 0.83 & 0.720 & conf & 0.048 & 0.8$\pm$0.2 & 10.0$\pm$0.8 & 4.2$\pm$1.0\\ 
1993 & 334.939 & -27.917 & 0.20 & 0.207 & conf & 0.070 & 0.7$\pm$0.2 & 0.9$\pm$0.2 & 2.1$\pm$0.6\\ 
1994 & 334.900 & -28.167 & 0.30 & \  & \  & 0.093 & 0.8$\pm$0.2 & 2.7$\pm$0.5 & 3.0$\pm$0.8\\ 
1995 & 334.966 & -28.175 & F & \  & \  & 0.019 & * & * & *\\ 
1999 & 150.655 & -8.148 & 0.49 & 0.500 & phot & 0.039 & 0.8$\pm$0.2 & 3.8$\pm$0.6 & 3.2$\pm$0.8\\ 
2002 & 359.900 & -32.187 & F & 0.480 & phot & 0.109 & 0.9$\pm$0.2 & 8.6$\pm$1.2 & 4.4$\pm$1.1\\ 
2003 & 326.523 & 4.383 & 0.52 & 0.530 & conf & 0.115 & 0.9$\pm$0.2 & 12.8$\pm$0.7 & 5.0$\pm$1.2\\ 
2005 & 191.013 & 16.866 & 0.54 & 0.560 & conf & 0.093 & 0.8$\pm$0.2 & 6.7$\pm$2.0 & 3.9$\pm$1.2\\ 
2006 & 197.843 & -5.781 & 0.18 & 0.172 & tent & 0.030 & 0.6$\pm$0.2 & 0.4$\pm$0.1 & 1.5$\pm$0.4\\ 
2012 & 188.998 & -33.883 & 0.22 & 0.082 & tent & 0.057 & 0.7$\pm$0.2 & 1.4$\pm$0.4 & 2.4$\pm$0.7\\ 
2020 & 214.847 & 6.643 & 0.56 & 0.570 & phot & 0.144 & 1.0$\pm$0.2 & 16.1$\pm$2.1 & 5.4$\pm$1.3\\ 
2021 & 214.973 & 6.568 & 0.58 & 0.560 & phot & 0.156 & 1.0$\pm$0.3 & 18.3$\pm$2.7 & 5.6$\pm$1.5\\ 
2022 & 215.001 & 6.581 & 0.58 & 0.570 & phot & 0.087 & 0.9$\pm$0.2 & 10.8$\pm$1.6 & 4.6$\pm$1.2\\ 
2023 & 163.898 & -4.990 & 0.58 & 0.610 & phot & 0.032 & 0.7$\pm$0.2 & 4.4$\pm$0.6 & 3.3$\pm$0.8\\ 
2025 & 163.796 & -5.071 & 0.66 & 0.680 & conf & 0.061 & 0.8$\pm$0.2 & 11.2$\pm$0.8 & 4.5$\pm$1.1\\ 
2031 & 54.656 & -35.690 & 0.20 & 0.185 & conf & 0.053 & 0.6$\pm$0.1 & 0.4$\pm$0.1 & 1.6$\pm$0.4\\ 
2045 & 175.063 & 2.941 & 0.20 & \  & \  & 0.022 & 0.6$\pm$0.2 & 0.6$\pm$0.2 & 1.7$\pm$0.6\\ 
2046 & 218.702 & 3.631 & 0.13 & 0.146 & conf & 0.083 & 0.6$\pm$0.2 & 0.4$\pm$0.1 & 1.6$\pm$0.4\\ 
2048 & 54.547 & -22.941 & 0.18 & 0.173 & phot & 0.154 & 0.8$\pm$0.2 & 1.5$\pm$0.4 & 2.5$\pm$0.7\\ 
2049 & 54.461 & -23.074 & 0.62 & \  & \  & 0.038 & 0.7$\pm$0.2 & 5.7$\pm$1.1 & 3.6$\pm$0.9\\ 
2057 & 187.696 & 11.189 & D & \  & \  & 0.022 & * & * & *\\ 
2062 & 338.836 & -25.962 & D & 1.393 & phot & 0.024 & ** & ** & **\\ 
2063 & 147.072 & -13.279 & 0.06 & \  & \  & 0.039 & 0.5$\pm$0.2 & 0.2$\pm$0.1 & 1.2$\pm$0.6\\ 
2078 & 32.608 & -39.494 & F & 0.306 & conf & 0.050 & 0.7$\pm$0.2 & 1.6$\pm$0.1 & 2.4$\pm$0.6\\ 
2079 & 32.556 & -39.549 & 0.17 & 0.166 & conf & 0.075 & 0.7$\pm$0.2 & 0.6$\pm$0.1 & 1.8$\pm$0.5\\ 
2093 & 335.812 & -1.661 & 0.32 & 0.297 & phot & 0.265 & 1.0$\pm$0.3 & 9.2$\pm$1.7 & 4.8$\pm$1.1\\ 
2094 & 200.323 & -11.741 & 0.55 & \  & \  & 0.029 & 0.7$\pm$0.2 & 3.4$\pm$0.4 & 3.0$\pm$0.7\\ 
2099 & 323.423 & -42.729 & 0.19 & \  & \  & 0.103 & 0.7$\pm$0.2 & 1.2$\pm$0.3 & 2.3$\pm$0.6\\ 
2100 & 323.395 & -42.902 & 0.31 & \  & \  & 0.040 & 0.7$\pm$0.2 & 1.1$\pm$0.2 & 2.1$\pm$0.6\\ 
2115 & 188.598 & 15.316 & 0.30 & 0.308 & phot & 0.048 & 0.7$\pm$0.2 & 1.9$\pm$0.6 & 2.6$\pm$0.8\\ 
2118 & 327.847 & -5.448 & 0.16 & 0.145 & conf & 0.135 & 0.7$\pm$0.2 & 0.8$\pm$0.1 & 2.0$\pm$0.5\\ 
2122 & 308.703 & -34.530 & 0.37 & \  & \  & 0.164 & 0.9$\pm$0.2 & 7.4$\pm$1.1 & 4.3$\pm$1.1\\ 
2128 & 157.532 & -3.111 & 0.45 & 0.430 & phot & 0.047 & 0.7$\pm$0.2 & 3.0$\pm$0.6 & 3.0$\pm$0.8\\ 
2130 & 329.308 & -7.712 & 0.47 & 0.450 & phot & 0.034 & 0.7$\pm$0.7 & 2.7$\pm$2.7 & 2.8$\pm$2.9\\ 
2161 & 34.009 & -47.876 & 0.59 & \  & \  & 0.011 & 0.6$\pm$0.1 & 1.3$\pm$0.5 & 2.1$\pm$0.5\\ 
2162 & 149.853 & 1.772 & 0.12 & 0.120 & conf & 0.079 & 0.6$\pm$0.2 & 0.3$\pm$0.1 & 1.4$\pm$0.4\\ 
2163 & 149.965 & 1.680 & 0.33 & 0.370 & conf & 0.056 & 0.8$\pm$0.2 & 2.8$\pm$0.2 & 3.0$\pm$0.8\\ 
2166 & 349.197 & -42.711 & 0.11 & 0.096 & conf & 0.278 & 0.7$\pm$0.2 & 0.7$\pm$0.1 & 1.9$\pm$0.5\\ 
2169 & 198.667 & -25.340 & 0.23 & 0.250 & tent & 0.189 & 0.9$\pm$0.3 & 3.8$\pm$1.1 & 3.5$\pm$1.1\\ 
2187 & 197.876 & -5.869 & 0.45 & 0.461 & conf & 0.086 & 0.9$\pm$0.2 & 6.3$\pm$0.4 & 3.9$\pm$0.9\\ 
2189 & 352.216 & 14.882 & 0.47 & 0.497 & conf & 0.044 & 0.8$\pm$0.2 & 4.3$\pm$0.2 & 3.3$\pm$0.8\\ 
2199 & 309.625 & -1.424 & 0.81 & 0.680 & conf & 0.051 & 0.8$\pm$0.2 & 11.2$\pm$1.3 & 4.5$\pm$1.1\\ 
2203 & 341.053 & -9.575 & 0.44 & 0.447 & conf & 0.184 & 1.0$\pm$0.2 & 13.0$\pm$1.4 & 5.2$\pm$1.2\\ 
2207 & 192.362 & 5.208 & 0.62 & \  & \  & 0.020 & 0.7$\pm$0.2 & 3.3$\pm$0.7 & 2.9$\pm$0.7\\ 
2209 & 149.769 & 13.089 & 0.36 & 0.396 & conf & 0.118 & 0.9$\pm$0.2 & 6.4$\pm$0.9 & 4.0$\pm$1.1\\ 
2212 & 189.708 & 9.254 & 0.80 & \  & \  & 0.042 & 0.8$\pm$0.2 & 12.2$\pm$3.0 & 4.5$\pm$1.2\\ 
2225 & 14.396 & -26.112 & 0.36 & \  & \  & 0.063 & 0.8$\pm$0.2 & 2.7$\pm$0.7 & 3.0$\pm$0.8\\ 
2254 & 38.264 & -71.275 & 0.55 & \  & \  & 0.172 & 1.0$\pm$0.3 & 16.7$\pm$2.4 & 5.5$\pm$1.4\\ 
2255 & 225.214 & -10.861 & 0.40 & \  & \  & 0.025 & 0.7$\pm$0.2 & 2.7$\pm$1.1 & 2.9$\pm$1.0\\
\hline
 \end{tabular}
\end{table*}
\begin{table*}
 \contcaption{The X-CLASS/GROND catalogue}
 \label{tab:continued}
 \begin{tabular}{cccccccccc}
  \hline
  ID & RA J2000& DEC J2000 & $z$  & $z_{lit}$ & $z$-Type & Count rate  & $r_{500}$ & $ L_{500}^{[0.5-2] keV}$ & $T_{L-T} $ \\ 
X-CLASS & (degrees) & (degrees) & GROND & Literature & Literature & (cts/s) & (Mpc) & ($10^{43}\mathrm{erg\ s^{-1}}$) & (keV)\\
(1) & (2) & (3) & (4) & (5) & (6) & (7) & (8) & (9) & (10) \\ \hline
2256 & 225.275 & -10.876 & 0.76 & \  & \  & 0.042 & 0.8$\pm$0.2 & 12.4$\pm$1.9 & 4.6$\pm$1.1\\ 
2257 & 334.149 & -36.799 & 0.57 & \  & \  & 0.033 & 0.7$\pm$0.2 & 4.4$\pm$0.7 & 3.3$\pm$0.8\\ 
2260 & 187.211 & 13.995 & 0.50 & \  & \  & 0.085 & 0.9$\pm$0.2 & 7.7$\pm$1.2 & 4.2$\pm$1.0\\ 
2265 & 343.444 & -14.208 & 0.32 & \  & \  & 0.034 & 0.7$\pm$0.2 & 1.3$\pm$0.3 & 2.2$\pm$0.6\\ 
2294 & 5.622 & 1.383 & 0.61 & 0.620 & tent & 0.038 & 0.7$\pm$0.2 & 5.7$\pm$0.8 & 3.6$\pm$0.9\\ 
2297 & 15.127 & -47.823 & 0.42 & \  & \  & 0.154 & 1.0$\pm$0.3 & 9.0$\pm$1.4 & 4.5$\pm$1.2\\ 
2298 & 15.239 & -47.860 & 0.28 & \  & \  & 0.062 & 0.7$\pm$0.2 & 1.0$\pm$0.4 & 2.0$\pm$0.7\\ 
2299 & 86.974 & -47.651 & 0.45 & \  & \  & 0.026 & 0.7$\pm$0.2 & 2.3$\pm$0.6 & 2.7$\pm$0.7\\ 
2303 & 73.126 & -42.153 & 0.73 & \  & \  & 0.029 & 0.7$\pm$0.2 & 6.3$\pm$1.4 & 3.6$\pm$0.8\\ 
2304 & 179.895 & -19.862 & D & \  & \  & 0.069 & * & * & *\\ 
2305 & 180.059 & -20.047 & 0.60 & \  & \  & 0.279 & 1.1$\pm$0.3 & 37.2$\pm$4.4 & 7.3$\pm$1.8\\ 
2307 & 29.323 & -16.991 & 0.50 & \  & \  & 0.032 & 0.7$\pm$0.2 & 3.0$\pm$0.6 & 2.9$\pm$0.8\\ 
2311 & 141.282 & 13.450 & F & \  & \  & 0.048 & * & * & *\\ 
2312 & 141.206 & 13.293 & D & 0.520 & phot & 0.031 & 0.7$\pm$0.2 & 3.4$\pm$0.5 & 3.0$\pm$0.8\\ 
2313 & 53.003 & -27.724 & D & \  & \  & 0.012 & * & * & *\\ 
2321 & 137.723 & -9.738 & 0.08 & 0.092 & tent & 0.080 & 0.5$\pm$0.1 & 0.2$\pm$0.1 & 1.1$\pm$0.4\\ 
2323 & 245.403 & -1.491 & 0.11 & 0.106 & tent & 0.049 & 0.5$\pm$0.1 & 0.2$\pm$0.1 & 1.2$\pm$0.4\\ 
3075 & 28.173 & -13.649 & D & 0.830 & conf & 0.032 & 0.7$\pm$0.2 & 9.4$\pm$0.9 & 4.0$\pm$0.9\\ 
3104 & 327.673 & -5.685 & 0.36 & 0.440 & conf & 0.045 & 0.8$\pm$0.2 & 3.1$\pm$0.3 & 3.0$\pm$0.7\\ 
3170 & 184.205 & -12.137 & 0.79 & 0.480 & phot & 0.014 & 0.6$\pm$0.2 & 4.5$\pm$0.8 & 3.1$\pm$0.8\\ 
3281 & 3.386 & -27.188 & 0.50 & \  & \  & 0.054 & 0.8$\pm$0.2 & 5.0$\pm$0.7 & 3.5$\pm$0.9\\ 
3283 & 146.378 & 9.776 & 0.21 & 0.220 & conf & 0.047 & 0.7$\pm$0.2 & 0.9$\pm$0.2 & 2.0$\pm$0.6\\ 
3485 & 351.361 & -12.068 & 0.08 & 0.085 & conf & 0.154 & ** & ** & **\\ 

\hline
 \end{tabular}
\end{table*}


\bsp	
\label{lastpage}
\end{document}